%% file: paper.tex
\pdfoutput=1
\documentclass[12pt,a4paper]{article}
\input{paper_format}

\input{definitions}

\begin{document}

\renewcommand{\thefootnote}{\fnsymbol{footnote}}
\setcounter{footnote}{1}
\input{title_page}

\clearpage

\renewcommand{\thefootnote}{\arabic{footnote}}
\setcounter{footnote}{0}
\pagestyle{plain} 
\setcounter{page}{1}
\pagenumbering{arabic}


\input{introduction}
\input{CPasymmetry}
\input{amplitudeRelations}
\input{TDCP_Expression}
\input{appliExpResults}
\input{prospects}
\input{conclusion}
\input{acknowledgements}
\clearpage
\input{appendix_S-KsPiz_S-Phi}

\bibliographystyle{bibstyle}
\bibliography{paper}

\end{document}

%% file: paper_format.tex
\usepackage[top=1in, bottom=1.25in, left=1in, right=1in]{geometry}

\usepackage{microtype}
\usepackage{lineno}  
\usepackage{xspace} 
\usepackage{caption} 

\usepackage{graphicx}  
\usepackage{color}
\usepackage{colortbl}
\graphicspath{{./figs/}} 

\usepackage{amsmath} 
\usepackage{amssymb}
\usepackage{amsfonts}
\usepackage{upgreek} 

\usepackage{empheq}
\usepackage{bigints}

\usepackage{relsize}

\newcommand*\patchAmsMathEnvironmentForLineno[1]{%
\expandafter\let\csname old#1\expandafter\endcsname\csname #1\endcsname
\expandafter\let\csname oldend#1\expandafter\endcsname\csname
end#1\endcsname
 \renewenvironment{#1}%
   {\linenomath\csname old#1\endcsname}%
   {\csname oldend#1\endcsname\endlinenomath}%
}
\newcommand*\patchBothAmsMathEnvironmentsForLineno[1]{%
  \patchAmsMathEnvironmentForLineno{#1}%
  \patchAmsMathEnvironmentForLineno{#1*}%
}
\AtBeginDocument{%
\patchBothAmsMathEnvironmentsForLineno{equation}%
\patchBothAmsMathEnvironmentsForLineno{align}%
\patchBothAmsMathEnvironmentsForLineno{flalign}%
\patchBothAmsMathEnvironmentsForLineno{alignat}%
\patchBothAmsMathEnvironmentsForLineno{gather}%
\patchBothAmsMathEnvironmentsForLineno{multline}%
\patchBothAmsMathEnvironmentsForLineno{eqnarray}%
}

\usepackage{hyperref}    
\usepackage[all]{hypcap} 

\usepackage{cite} 
\usepackage{mciteplus}

\usepackage{epstopdf}
\usepackage{verbatim}
\usepackage[dvipsnames]{xcolor}
\RequirePackage{xspace}

\usepackage{longtable}
\usepackage{multirow}

%% file: definitions.tex
\def\es#1{\textcolor{black}{#1}}
\def\sa#1{\textcolor{black}{#1}}
\def\sas#1{\textcolor{black}{#1}}

\def\MomNormal  {\ensuremath{{\color{black}{p_1,p_2}},p_3}\xspace}
\def\MomSwapped {\ensuremath{{\color{black}{p_2,p_1}},p_3}\xspace}
\def\OrderNormal  {\ensuremath{{\color{black}{12}}3}\xspace}
\def\OrderSwapped  {\ensuremath{{\color{black}{21}}3}\xspace}

\newcommand{\figref}[1]{Fig.~\ref{#1}}

\newcommand{\secref}[1]{Sec.~\ref{#1}}

\newcommand{\Eqref}[1]{Eq.~(\ref{#1})}
\newcommand{\Eqsref}[2]{Eqs.~(\ref{#1}) and~(\ref{#2})}

\newcommand{\Mod}[1]   {\ensuremath{\left|{#1}\right|}\xspace}
\newcommand{\ModSq}[1] {\ensuremath{\left|{#1}\right|^2}\xspace}
\newcommand{\Real}[1]  {\ensuremath{\mathrm{Re} \left( {#1} \right)}\xspace}
\newcommand{\Imag}[1]  {\ensuremath{\mathrm{Im} \left( {#1} \right)}\xspace}

\newcommand{\ket}[1]   {\ensuremath{| {#1} \rangle}\xspace}
\newcommand{\braket}[3]{\ensuremath{\langle {#1} | {#2} | {#3} \rangle}\xspace}

\newlength{\textlarg}

 \def\PJ      {\ensuremath{J}\xspace}

 \def\Pb      {\ensuremath{b}\xspace}                 
                  
 \def\Pd      {\ensuremath{d}\xspace}

 \def\Pi      {\ensuremath{i}\xspace}

 \def\Ps      {\ensuremath{s}\xspace}                 
 \def\Pt      {\ensuremath{t}\xspace}

\def\dquark    {\ensuremath{\Pd}\xspace}

\def\squark    {\ensuremath{\Ps}\xspace}
\def\squarkbar {\ensuremath{\overline \squark}\xspace}

\def\bquark    {\ensuremath{\Pb}\xspace}
\def\bquarkbar {\ensuremath{\overline \bquark}\xspace}

\def\tquark    {\ensuremath{\Pt}\xspace}

\def\fb        {\ensuremath{\, \mathrm{fb}^{-1}}\xspace}
\def\ab        {\ensuremath{\, \mathrm{ab}^{-1}}\xspace}

\def\ms        {\ensuremath{m_{\squark}}\xspace}
\def\mb        {\ensuremath{m_{\bquark}}\xspace}

\def\jpsi      {\ensuremath{\PJ\mskip -3mu/\mskip -2mu\psi\mskip 2mu}\xspace}

\def\Vtd       {\ensuremath{V_{\tquark\dquark}}\xspace}

\def\Vts       {\ensuremath{V_{\tquark\squark}}\xspace}

\def\Vtb       {\ensuremath{V_{\tquark\bquark}}\xspace}

\def\Vtdstar   {\ensuremath{V_{\tquark\dquark}^{\conj}}\xspace}

\def\Vtsstar   {\ensuremath{V_{\tquark\squark}^{\conj}}\xspace}

\def\Vtbstar   {\ensuremath{V_{\tquark\bquark}^{\conj}}\xspace}

\def\text#1{\rm{#1}}

\def\Bbar    {{\ensuremath{\kern 0.18em\overline{\kern -0.18em B}{}}}\xspace}
\def\Bb      {{\ensuremath{\Bbar}}\xspace}
\def\BorBbar {\kern 0.18em\optbar{\kern -0.18em B}{}\xspace}

\def\B       {{\ensuremath{B}}\xspace}
\def\Bz      {{\ensuremath{B^0}}\xspace}
\def\Bzb     {{\ensuremath{\Bbar{}^0}}\xspace}
\def\Bu      {{\ensuremath{B^{+}}}\xspace}

\def\Bp      {{\ensuremath{\Bu}}\xspace}

\def\Bs      {{\ensuremath{B^0_s}}\xspace}
\def\Bsb     {{\ensuremath{\Bbar{}^0_s}}\xspace}

\def\pip     {{\ensuremath{\pi^+}}\xspace}
\def\pim     {{\ensuremath{\pi^-}}\xspace}

\def\piz     {{\ensuremath{\pi^0}}\xspace}
\def\rhoz    {{\ensuremath{\rho^0}}\xspace}
\def\Kbar    {{\ensuremath{\overline{K}}}\xspace}

\def\Kp      {{\ensuremath{K^{+}}}\xspace}
\def\Km      {{\ensuremath{K^{-}}}\xspace}

\def\Kstarz  {{\ensuremath{K^{*0}(892)}}\xspace}

\def\Kst     {{\ensuremath{K^{*}}}\xspace}
\def\Kstb    {{\ensuremath{\Kbar^{*}}}\xspace}
\def\Kstz    {{\ensuremath{K^{*0}}}\xspace}
\def\Kstzb    {{\ensuremath{\Kbar^{*0}}}\xspace}
\def\Kstp    {{\ensuremath{K^{*+}}}\xspace}
\def\Kstm    {{\ensuremath{K^{*-}}}\xspace}
\def\Kstpm   {{\ensuremath{K^{*\pm}}}\xspace}
\def\KS      {{\ensuremath{K^0_{\scriptscriptstyle S}}\xspace}}
\def\Kres    {\ensuremath{K_{\rm res}}\xspace}
\def\Kresbar {{\ensuremath{\overline{K}_{\rm res}}}\xspace}
\def\Kresp   {{\ensuremath{K^{+}_{\rm res}}}\xspace}

\def\swave   {{\ensuremath{(K\pi)_0}}\xspace} 
\def\swavep  {{\ensuremath{(K\pi)_0^+}}\xspace} 
\def\swavem  {{\ensuremath{(K\pi)_0^-}}\xspace}

\def\g           {\ensuremath{\gamma}\xspace}
\def\gL          {\ensuremath{\g_{L}}\xspace}
\def\gR          {\ensuremath{\g_{R}}\xspace}

\def\rhoKs       {\ensuremath{\rhoz \KS}\xspace}
\def\KstPi       {\ensuremath{\Kstp\pim}\xspace}
\def\KstPibar    {\ensuremath{\Kstm\pip}\xspace}
\def\KappaPi     {\ensuremath{\swavep\pim}\xspace}
\def\KappaPibar  {\ensuremath{\swavem\pip}\xspace}

\def\btodg    {{\ensuremath{\bquark \to \dquark \g}}\xspace}
\def\btosg    {{\ensuremath{\bquark \to \squark \g}}\xspace}
\def\btosgbar {{\ensuremath{\bquarkbar \to \squarkbar \g}}\xspace}

\def\dalitz         {\ensuremath{{p}}\xspace}

\def\SP {{\ensuremath{\mathcal{S}^+}}\xspace} 
\def\SM {{\ensuremath{\mathcal{S}^-}}\xspace} 

\def\SKspiz {{\ensuremath{\mathcal{S}_{\piz \KS \g}}}\xspace} 
\def\CKspiz {{\ensuremath{\mathcal{C}_{\piz \KS \g}}}\xspace} 

\def\SKspipi {{\ensuremath{\mathcal{S}_{\pip \pim \KS \g}}}\xspace} 
\def\CKspipi {{\ensuremath{\mathcal{C}_{\pip \pim \KS \g}}}\xspace} 
\def\SKspipiDalitz {{\ensuremath{\mathcal{S}^{\delta\dalitz}_{\pip \pim \KS \g}}}\xspace} 
 
\def\SKspipiDalitzI {{\ensuremath{\mathcal{S}^I_{\pip \pim \KS \g}}}\xspace}

\def\SKspipiDalitzIb {{\ensuremath{\mathcal{S}^{\overline{I}}_{\pip \pim \KS \g}}}\xspace}

\def\SKsrho  {{\ensuremath{\mathcal{S}_{\rhoKs \g}}}\xspace} 
\def\CKsrho  {{\ensuremath{\mathcal{C}_{\rhoKs \g}}}\xspace} 
\def\mKpipi  {{\ensuremath{m_{K  \pi \pi}}}\xspace}

\def\CP      {{\ensuremath{C\!P}}\xspace}

\def\D       {\ensuremath{{\mathcal D}}\xspace}

\def \nChannelRhoz  {\mbox{{\ensuremath{B^0 \to \rhoz \KS \g}}}\xspace}

\newcommand{\mevcc}{\ensuremath{{\mathrm{\,Me\kern -0.1em V\!/}c^2}}\xspace}
\newcommand{\gevcc}{\ensuremath{{\mathrm{\,Ge\kern -0.1em V\!/}c^2}}\xspace}
\newcommand{\gevcccc}{\ensuremath{{\mathrm{\,Ge\kern -0.1em V^2\!/}c^4}}\xspace}

\usepackage{relsize}
\def\babar{\mbox{\slshape B\kern-0.1em{\smaller A}\kern-0.1em
    B\kern-0.1em{\smaller A\kern-0.2em R}}\xspace}
\def\superbelle	{\mbox{{\ensuremath{{\rm Belle~II}}}}\xspace}

\def\conj         {\ensuremath{*}\xspace}

\def\qoverp       {\ensuremath{{\frac{q}{p}}}\xspace}

\def\CS    		  	{\ensuremath{C_7}\xspace}
\def\CSp      		{\ensuremath{C^{\prime}_7}\xspace}
\def\CSpoverCS      {\ensuremath{\CSp/\CS}\xspace}

\def\Wc              {\ensuremath{c}\xspace}

\def\cstar             {\ensuremath{\Wc^{\conj}}\xspace}
\def\cprime            {\ensuremath{\Wc^{\prime}}\xspace}
\def\cprimestar        {\ensuremath{\Wc^{\prime \conj}}\xspace}

\def\covercprime           {\ensuremath{{\frac{\Wc}{\cprime}}}\xspace}

\def\covercprimestar       {\ensuremath{{\frac{\Wc}{\cprimestar}}}\xspace}

\def\cprimeoverc           {\ensuremath{{\frac{\cprime}{\Wc}}}\xspace}
\def\cprimeovercstar       {\ensuremath{{\frac{\cprime}{\cstar}}}\xspace}

\def\CprimeoverCprimestar  {\ensuremath{{\frac{\cprime}{\cprimestar}}}\xspace}

\def\cstarovercprimestar   {\ensuremath{{\frac{\cstar}{\cprimestar}}}\xspace}

\def\dm            {\ensuremath{\Delta m_{}}\xspace}

\def\H             {\ensuremath{{\mathcal H}}\xspace}
\def\Hp            {\ensuremath{\H^{+}}\xspace}
\def\Hpdag         {\ensuremath{\H^{+\dagger}}\xspace}
\def\Hm            {\ensuremath{\H^{-}}\xspace}
\def\Hmdag         {\ensuremath{\H^{-\dagger}}\xspace}
\def\Hs            {\ensuremath{\H_s}\xspace}
\def\Hsprime       {\ensuremath{\H_s^{\prime}}\xspace}

\def\C            {\ensuremath{{\mathcal C}}\xspace}
\def\Cdag         {\ensuremath{\C^{\dagger}}\xspace}
\def\P            {\ensuremath{{\mathcal P}}\xspace}
\def\Pdag         {\ensuremath{\P^{\dagger}}\xspace}

\newcommand{\sumLR}[1]  {\ensuremath{\sum\limits_{\lambda={L,R}} \left[ {#1} \right]}\xspace}

\def\A              {\ensuremath{{M}}\xspace}
\def\Abar           {\ensuremath{\overline{\A}}\xspace}

\def\AmpTot         {\ensuremath{\A_{\lambda}}\xspace}
\def\AmpTotL        {\ensuremath{\A_{L}}\xspace}
\def\AmpTotR        {\ensuremath{\A_{R}}\xspace}
\def\AmpTotBar      {\ensuremath{\Abar_{\lambda}}\xspace}
\def\AmpTotBarL     {\ensuremath{\Abar_{L}}\xspace}
\def\AmpTotBarR     {\ensuremath{\Abar_{R}}\xspace}

\def\AmpTotConj     {\ensuremath{\AmpTot^{\conj}}\xspace}
\def\AmpTotLConj    {\ensuremath{\AmpTotL^{\conj}}\xspace}
\def\AmpTotRConj    {\ensuremath{\AmpTotR^{\conj}}\xspace}

\def\AmpTotRhoKs      {\ensuremath{\AmpTot^{\rhoKs}}\xspace}
\def\AmpTotLRhoKs     {\ensuremath{\AmpTotL^{\rhoKs}}\xspace}

\def\AmpTotBarRhoKs   {\ensuremath{\AmpTotBar^{\rhoKs}}\xspace}

\def\AmpTotRhoKsConj      {\ensuremath{\AmpTot^{\conj\rhoKs}}\xspace}
\def\AmpTotLRhoKsConj     {\ensuremath{\AmpTotL^{\conj\rhoKs}}\xspace}

\def\AmpTotBarRhoKsConj   {\ensuremath{\AmpTotBar^{\conj\rhoKs}}\xspace}

\def\AmpTotKstPi      {\ensuremath{\AmpTot^{\KstPi}}\xspace}
\def\AmpTotLKstPi     {\ensuremath{\AmpTotL^{\KstPi}}\xspace}

\def\AmpTotBarKstPi   {\ensuremath{\AmpTotBar^{\KstPibar}}\xspace}

\def\AmpTotKstPiConj      {\ensuremath{\AmpTot^{\conj\KstPi}}\xspace}

\def\AmpTotBarKstPiConj   {\ensuremath{\AmpTotBar^{\conj\KstPibar}}\xspace}

\def\AmpTotKappaPi      {\ensuremath{\AmpTot^{\KappaPi}}\xspace}
\def\AmpTotLKappaPi     {\ensuremath{\AmpTotL^{\KappaPi}}\xspace}

\def\AmpTotBarKappaPi   {\ensuremath{\AmpTotBar^{\KappaPibar}}\xspace}

\def\AmpTotKappaPiConj      {\ensuremath{\AmpTot^{\conj\KappaPi}}\xspace}

\def\AmpTotLIConj     {\ensuremath{\AmpTotL^ {\ensuremath{\conj i}\xspace}}\xspace}
\def\AmpTotLJ      {\ensuremath{\AmpTotL^{\ensuremath{j}\xspace}}\xspace}

\def\AW               {\ensuremath{A}\xspace}
\def\AWBar            {\ensuremath{\overline{\AW}}\xspace}

\def\AmpWeak          {\ensuremath{\AW_{\lambda}}\xspace}
\def\AmpWeakL         {\ensuremath{\AW_{L}}\xspace}
\def\AmpWeakR         {\ensuremath{\AW_{R}}\xspace}
\def\AmpWeakLR        {\ensuremath{\AW_{L,R}}\xspace}
\def\AmpWeakBar       {\ensuremath{\AWBar_{\lambda}}\xspace}
\def\AmpWeakLBar      {\ensuremath{\AWBar_{L}}\xspace}
\def\AmpWeakRBar      {\ensuremath{\AWBar_{R}}\xspace}
\def\AmpWeakLRBar     {\ensuremath{\AWBar_{L,R}}\xspace}

\def\AS                 {\ensuremath{\mathcal{A}}\xspace}
\def\ASBar              {\ensuremath{\overline{\AS}}\xspace}

\def\AmpStrong          {\ensuremath{\AS_{\lambda}}\xspace}
\def\AmpStrongL         {\ensuremath{\AS_{L}}\xspace}
\def\AmpStrongR         {\ensuremath{\AS_{R}}\xspace}

\def\AmpStrongBar       {\ensuremath{\ASBar_{\lambda}}\xspace}
\def\AmpStrongLBar      {\ensuremath{\ASBar_{L}}\xspace}
\def\AmpStrongRBar      {\ensuremath{\ASBar_{R}}\xspace}

\def\AmpStrongGeneric      {\ensuremath{\AmpStrong^{\,i}}\xspace}
\def\AmpStrongLGeneric     {\ensuremath{\AmpStrongL^{\,i}}\xspace}
\def\AmpStrongRGeneric     {\ensuremath{\AmpStrongR^{\,i}}\xspace}

\def\AmpStrongBarGeneric   {\ensuremath{\AmpStrongBar^{\,i}}\xspace}
\def\AmpStrongLBarGeneric  {\ensuremath{\AmpStrongLBar^{\,i}}\xspace}
\def\AmpStrongRBarGeneric  {\ensuremath{\AmpStrongRBar^{\,i}}\xspace}

\def\AmpStrongLGenericConj     {\ensuremath{\AmpStrongL^{\conj i}}\xspace}
\def\AmpStrongRGenericConj     {\ensuremath{\AmpStrongR^{\conj i}}\xspace}

\def\AmpStrongLBarGenericConj  {\ensuremath{\AmpStrongLBar^{\conj i}}\xspace}
\def\AmpStrongRBarGenericConj  {\ensuremath{\AmpStrongRBar^{\conj i}}\xspace}

\def\AmpStrongRhoKs      {\ensuremath{\AmpStrong^{\rhoKs}}\xspace}

\def\AmpStrongBarRhoKs   {\ensuremath{\AmpStrongBar^{\rhoKs}}\xspace}

\def\AmpStrongKstPi      {\ensuremath{\AmpStrong^{\KstPi}}\xspace}

\def\AmpStrongBarKstPi   {\ensuremath{\AmpStrongBar^{\KstPibar}}\xspace}

\def\AmpStrongKappaPi      {\ensuremath{\AmpStrong^{\KappaPi}}\xspace}

\def\AmpStrongBarKappaPi   {\ensuremath{\AmpStrongBar^{\KappaPibar}}\xspace}

\def\AmpStrongG      {\ensuremath{\AS^{\xspace}}\xspace}
\def\AmpStrongGConj     {\ensuremath{\AS^ {\ensuremath{\conj}\xspace}}\xspace}
\def\AmpStrongG      {\ensuremath{\AS^{\xspace}}\xspace}
\def\AmpStrongGConj     {\ensuremath{\AS^ {\ensuremath{\conj}\xspace}}\xspace}

\def\AmpStrongGI      {\ensuremath{\AS^{\ensuremath{i}\xspace}}\xspace}
\def\AmpStrongGIConj     {\ensuremath{\AS^ {\ensuremath{\conj i}\xspace}}\xspace}
\def\AmpStrongGJ      {\ensuremath{\AS^{\ensuremath{j}\xspace}}\xspace}

\def\AmpStrongGRhoKs      {\ensuremath{\AS^{\ensuremath{\rhoKs}\xspace}}\xspace}
\def\AmpStrongGKstPi      {\ensuremath{\AS^{\ensuremath{\KstPi}\xspace}}\xspace}
\def\AmpStrongGKappaPi      {\ensuremath{\AS^{\ensuremath{\KappaPi}\xspace}}\xspace}

\def\AmpStrongGRhoKsConj      {\ensuremath{\AS^{\ensuremath{\conj \rhoKs}\xspace}}\xspace}
\def\AmpStrongGKstPiConj      {\ensuremath{\AS^{\ensuremath{\conj \KstPi}\xspace}}\xspace}
\def\AmpStrongGKappaPiConj      {\ensuremath{\AS^{\ensuremath{\conj \KappaPi}\xspace}}\xspace}

\def\ASDalitzNormal {\ensuremath{\AS^{}_{\OrderNormal }}}
\def\ASDalitzConjNormal {\ensuremath{\AS^{\conj}_{\OrderNormal }}}
\def\ASDalitzSwapped {\ensuremath{\AS^{}_{\OrderSwapped }}}

%% file: title_page.tex
\begin{titlepage}
\pagenumbering{roman}

\vspace*{-1.5cm}
\rightline{September 5, 2019}
\vspace*{2.0cm}

{\normalfont\bfseries\boldmath\LARGE
\begin{center}
The time-dependent $C\!P$ asymmetry in $B^0 \to  K_{\rm res} \gamma \to \pi^+ \pi^- K^0_{\scriptscriptstyle S} \gamma$ decays
\end{center}
}
\vspace*{2.0cm}

\begin{center}
\sc{ \normalsize $^{a,b}$S. Akar, $^b$E. Ben-Haim, $^c$J. Hebinger, $^c$E. Kou and $^d$F.-S. Yu} \\
\vspace{1cm}
{\it\footnotesize $^a$University of Cincinnati, Cincinnati, OH, United States}\\[5pt]

{\it\footnotesize $^b$LPNHE, Sorbonne Universit\'e, Paris Diderot Sorbonne Paris Cit\'e, CNRS/IN2P3, Paris, France}\\[6pt]

{\it\footnotesize $^c$LAL,
  Universit\'e Paris-Sud, CNRS/IN2P3, Universit\'e Paris-Saclay, Orsay, France}\\[3pt]

{\it\footnotesize $^d$School of Nuclear Science and Technology, Lanzhou University, Lanzhou 730000, China}
\end{center}

\vspace{\fill}

\begin{abstract}
\noindent

The time-dependent $C\!P$ asymmetry in \mbox{$B^0 \to  K_{\rm res} \gamma  \to \pi^+ \pi^- K^0_{\scriptscriptstyle S} \gamma$} is sensitive to the photon polarisation in the quark level process $b \to s \gamma$. 
While this polarisation is predominantly left-handed in the standard model, it could be modified by the existence of new physics contributions that may possess different $C\!P$ properties. 
In this paper, we derive the $C\!P$ violation formulae for $B^0 \to K_{\rm res} \gamma  \to \pi^+ \pi^- K^0_{\scriptscriptstyle S} \gamma$ including the most dominant intermediate states. 
We propose a new observable that could be measured in a time-dependent amplitude analysis of $B^0 \to \pi^+ \pi^- K^0_{\scriptscriptstyle S} \gamma$ decays, providing a stringent contraint on the photon polarisation.
We discuss the future prospects for obtaining such constraints from measurements at Belle~II and LHCb.

\end{abstract}

\vspace*{1.0cm}

\begin{center}
  Published in JHEP 09 (2019) 034
\end{center}

\vspace{\fill}
\end{titlepage}


\newpage
\setcounter{page}{2}
\mbox{~}

\cleardoublepage

%% file: introduction.tex
\section{Introduction}
\label{sec:introduction}

The exclusive \btosg process is one of the most sensitive observables to new physics in \B physics: unlike many other $b$-hadron decays, it is described in the standard model (SM) by a single operator, the electro-magnetic type \mbox{$\squarkbar \sigma_{\mu\nu} (1+\g_5) \bquark F^{\mu\nu}$},  which minimises the uncertainties from hadronic effects. 
In the era of the LHC and the upgraded \B-factory experiment, \superbelle, an interesting opportunity opens to investigate the circular-polarisation of the photon in \btosg process and gain additional insight into its nature. 
In the standard model, the photon polarisation of \btosg is predicted to be predominantly left-handed ($\g_L$) due to the operator mentioned above.
Several new-physics  models contain new particles that couple differently from the SM, inducing an opposite chirality operator \mbox{$\squarkbar \sigma_{\mu\nu} (1-\g_5) \bquark F^{\mu\nu}$}; these models predict an enhanced right-handed photon contribution ($\g_R$). Examples of such new physics models are given in Refs.~\cite{Becirevic:2012dx,Kou:2013gna,Haba:2015gwa,Paul:2016urs}. 
The photon polarisation in \btosg transitions is therefore a fundamental property of the SM, and its experimental determination may provide information on physics beyond the SM.

Photon polarisation measurement is a challenge in \B physics, and much effort has been put into it in recent years.
Two types of methods to determine photon polarisation have been proposed and carried out: measuring the angular distribution of the recoil particles (\sa{see~\cite{Gronau:2001ng,Gronau:2002rz,Kou:2010kn, Kou:2016iau,Bishara:2015yta,Oliver:2010im,Kruger:2005ep,Becirevic:2011bp,Mannel:1997xy} for theoretical proposals and~\cite{Aaij:2015dea,Aaij:2014wgo,Aaij:2019hhx} for experimental results}), and measuring the time-dependent \CP asymmetry (Refs.~\cite{Atwood:1997zr,Atwood:2004jj,Atwood:2007qh, Muheim:2008vu} and~\cite{Aubert:2008gy,Ushiroda:2006fi, Li:2008qma,Sanchez:2015pxu,Aaij:2019pnd} for theory and experiment, respectively). 
In this article, we discuss the second method.

\begin{figure}[b]
{\relsize{-1.3}
\begin{minipage}{0.48\textwidth}
\begin{eqnarray}
 & \Bz \ {{\cprime(\Wc)}  \atop \longrightarrow}\ \Kres \g_{L (R)}  & \nonumber \\
\phantom{B(0)}^{ f_+(t)}\nearrow& \hspace*{2.cm}& \searrow \nonumber \\
\Bz(t=0)\hspace*{0.5cm}&\hspace*{2.cm}&  (n\pi) \KS \g_{L (R)} \; \nonumber \\
\phantom{B(0)}_{ \qoverp f_-(t)}\searrow & \hspace*{2.cm}& \nearrow\nonumber \\
& \Bzb \ {{\Wc(\cprime)}  \atop \longrightarrow}\ \Kresbar \g_{L (R)} & \nonumber \end{eqnarray}
\end{minipage}
\hspace*{0.3cm}
\begin{minipage}{0.48\textwidth}
\begin{eqnarray}
 & \Bz\ {{\cprime(\Wc)}  \atop \longrightarrow}\ \Kres \g_{L (R)}  & \nonumber \\
\phantom{B(0)}^{\frac{p}{q}f_-(t)}\nearrow& \hspace*{2.cm}& \searrow\nonumber \\
\Bzb(t=0)\hspace*{0.5cm}&\hspace*{2.cm}&   (n\pi) \KS \g_{L (R)} 
\nonumber \\ 
\phantom{B(0)}_{ f_+(t) }\searrow & \hspace*{2.cm}& \nearrow \nonumber \\
& \Bzb \ {{\Wc(\cprime)}  \atop \longrightarrow}\ \Kresbar \g_{L (R)} & \nonumber 
\end{eqnarray}
\end{minipage}
}
\caption{Schematic description of the time-dependent \CP asymmetry of \mbox{$B\to \Kres \g \to (n\pi)\KS \g$}.
The factors $f_-(t)$ and $f_+(t)$ are the time-dependent oscillation and non-oscillation probabilities, respectively, of a $\Bz$ or a $\Bzb$ meson.
The $q$ and $p$ are the $\B-\Bb$ oscillation parameters, which correspond to  \es{\mbox{$q/p \simeq(\Vtbstar\Vtd)/(\Vtb\Vtdstar)=e^{-2i\beta}$}} in the SM.
The coefficients \Wc and \cprime represent the ratio of the standard operator contribution \mbox{$\squarkbar \sigma_{\mu\nu} (1+\g_5) \bquark F^{\mu\nu}$} and that of the non-standard one \mbox{$\squarkbar \sigma_{\mu\nu} (1-\g_5) \bquark F^{\mu\nu}$}, respectively. 
In the SM, \es{$\cprime/\Wc\simeq\ms/\mb (\simeq 0)$} leading to an expected \CP asymmetry to be almost zero.
}
\label{fig:1}
 \end{figure}

Obtaining information on photon polarisation from the time-dependent \CP asymmetry measurement is illustrated with the promising mode \mbox{$\B \to \Kres \g \to (n\pi)\KS\g$}, where \Kres is a kaonic resonance \sa{and $n\pi$ designates either $\piz$ or $\pip\pim$}.
The illustration is depicted in Fig~\ref{fig:1}.
The time-dependent \CP asymmetry originates from the interference of the \mbox{$\B \to \Kres \g \to (n\pi) \KS \g$} and \mbox{$\Bb \to \Kresbar\g \to (n\pi)\KS\g$} amplitudes, one of which emerges as a result of \B-\Bb oscillation. 
The key point is that interference occurs only when photons coming from \B and \Bb amplitudes are circularly polarised in the same direction. 
Let us define the rate of  \Bb (\B) mesons decaying into left- (right-) handed photons to be \Wc, and the rate of \Bb (\B) into right- (left-) handed photons  to be \cprime. The former process is induced by the standard operator contribution \mbox{$\squarkbar \sigma_{\mu\nu} (1+\g_5) \bquark F^{\mu\nu}$}, and the latter is induced by the non-standard operator \mbox{$\squarkbar \sigma_{\mu\nu} (1-\g_5) \bquark F^{\mu\nu}$}.
Knowing that in the SM \es{\mbox{$\cprime/\Wc \simeq \ms/\mb \simeq 0$}}, that is, the left- (right-) handed photon is nearly forbidden for a \B (\Bb) meson decay, the interference of \B and \Bb is expected to be nearly zero. 
Therefore, observation of non-zero \CP asymmetry signals new physics. 
Once non-zero \CP asymmetry is observed, one can further determine the ``photon polarisation'', by measuring the ratio of $\cprime/\Wc$ using as input the oscillation parameters \sa{$q/p \simeq(\Vtbstar\Vtd)/(\Vtb\Vtdstar)=e^{-2i\beta}$ in the SM}. 
In this article, the notation used for the weak \Bz-\Bzb mixing phase is $\beta$ rather than its equivalent, $\phi_1$.
\sas{Its} numerical value is obtained from $\sin 2\beta$ measurements determined in other modes~\cite{Amhis:2016xyh}, such as \mbox{$\Bz \to \jpsi \KS$}.

The simplest decay mode to study in this regard is  \mbox{$\Bz \to \Kstarz \g \to \KS \piz \g$}, for which the first
measurements of the mixing-induced \CP violation were
reported by the \babar~\cite{Aubert:2008gy} and Belle~\cite{Ushiroda:2006fi} experiments:
\mbox{$S_{\KS\piz \g}^{\scriptsize}=-0.03\pm 0.29\pm 0.03$} and
\mbox{$S_{\KS\piz \g}^{}=-0.32^{+0.36}_{-0.33}\pm 0.05$}, respectively. 
As these measurements are statistically limited, they can be significantly improved by the \superbelle experiment, which plans to  accumulate a data sample 50 times as large as those accumulated by the first-generation \B factories.

In this paper, we discuss two methods to obtain information on the photon polarisation via the measurement of the mixing-induced \CP violation in the decay \mbox{$\Bz \to \Kres \g \to \rhoz \KS \g \to \pip \pim \KS \g$}. 
The main difficulty comes from the fact that the final state \mbox{$\pip \pim \KS$} can originate not only from the \CP eigenstate \rhoKs but also from other intermediate states. 
In order to disentangle these contributions, such as $K^{*\pm} \pi^\mp$, a detailed amplitude analysis is required. 
Such an analysis has been pioneered by the Belle collaboration~\cite{Li:2008qma} and extended by the \babar collaboration~\cite{Akar:2013ima}.

In this paper, motivated by these developments, we re-visit the method to obtain the mixing-induced \CP asymmetry in \mbox{$\Bz \to \Kres \g \to \pip \pim \KS \g$} decays to gain more insight on the photon polarisation.
\es{One essential ingredient of the method is the way \sa{to relate the} \B and \Bbar decay amplitudes for left- and right-handed photons in the final state, considering the dominant intermediate decay modes} 
\begin{eqnarray}
K_1(1270), \quad K_1(1400) && (J^P=1^+), \nonumber \\
\Kst(1410), \quad \Kst(1680) && (J^P=1^-), \nonumber \\
K_2^*(1430) && (J^P=2^+), \nonumber 
\end{eqnarray}
for the kaonic resonances and 
 \begin{eqnarray}
& \Bz \to \Kres \g \to (\rhoKs)   \g \to \KS(\pip\pim)\g,   \nonumber \\
& \Bz \to \Kres \g \to (\KstPi)   \g \to (\KS\pip)\pim\g, & \nonumber \\
& \Bz \to \Kres \g \to (\KappaPi) \g \to (\KS\pip)\pim\g, & \nonumber 
\end{eqnarray}
for the $K\pi$ or $\pi\pi$ intermediate states. The notation \swave designates the $K\pi$ $S$-wave.
\es{In practice the relation between amplitudes is obtained by studying the transformation of each intermediate state by parity (\P) and charge conjugation ($\mathcal{C}$).}

In \secref{sec:overview} we introduce the time-dependent \CP asymmetry formulae for \mbox{$\Bz \to \Kres \g \to \pip \pim \KS \g$} decays. 
In \secref{sec:amplitudes} we derive the \CP-sign for the decay amplitudes with different intermediate states, which is required in order to extract the \CP asymmetry.
Using these results we derive the time-dependent \CP asymmetry expression for \mbox{$\Bz \to \Kres \g \to \pip \pim \KS \g$} and \sa{\mbox{$\Bz \to \Kres \g \to \rhoz \KS \g$}} decays in \secref{sec:TDCP_Expression}.
In \secref{sec:application} we present two methods to obtain information on the photon polarisation.
Finally, in \secref{sec:interpretation} we discuss the future prospects for these measurements at \superbelle, and we conclude in \secref{sec:conclusion}.

%% file: CPasymmetry.tex
\section{\boldmath{Time-dependent \CP asymmetry for $\Bz \to \pip \pim \KS \g$}}\label{sec:overview}

In the limit where the rate ${\Gamma}_{\pip \pim \KS \g}(t)$ comes only from the amplitude for \sa{\mbox{$\Bz \to \Kres \g \to \rhoz \KS \g$}} decays, we define the time-dependent \CP asymmetry as
\begin{equation}
\frac{\overline{\Gamma}_{\rhoKs \g}(t)-\Gamma_{\rhoKs \g}(t)}{\overline{\Gamma}_{\rhoKs \g}(t)+\Gamma_{\rhoKs \g}(t)}\equiv 
\SKsrho \sin(\dm t) - \CKsrho \cos(\dm t),
\label{eq:2-1_v2}
\end{equation}
with
\begin{eqnarray}
\SKsrho &=& \frac{ 2\Imag{\mathlarger{\qoverp} \sa{\mathlarger{\int}} \sumLR{\AmpTotRhoKsConj \AmpTotBarRhoKs} \sa{dp}}}{\sa{\mathlarger{\int}}\sumLR{ \ModSq{\AmpTotBarRhoKs} + \ModSq{\AmpTotRhoKs}} \sa{dp}}, \label{eq:v3-2_v2a}
\\ [0.5em]
\CKsrho &=&-\frac{\sa{\mathlarger{\int}} \sumLR{ \ModSq{\AmpTotBarRhoKs} - \ModSq{\AmpTotRhoKs}} \sa{dp}}{\sa{\mathlarger{\int}} \sumLR{ \ModSq{\AmpTotBarRhoKs} + \ModSq{\AmpTotRhoKs}} \sa{dp}}, \label{eq:v3-2_v2b}
\end{eqnarray}
where \AmpTotRhoKs and \AmpTotBarRhoKs correspond to the \sa{$\Bz \to \Kres \g \to \rhoz \KS \g$} and \sa{$\Bzb \to \Kresbar \g \to \rhoz \KS \g$} decay amplitudes, respectively, with the left- and right-handed photon polarisation, designated by $\lambda=L, R$.
\sa{For simplicity, we first consider the contribution of a single kaonic resonance during the development of the formalism used in this article.} 
\sa{This simplification is justified in \secref{sec:TDCP_Expression} where it is shown that, depending on the considered phase-space region, the mixing-induced \CP-violation parameter expression does not depend on the kaonic resonance quantum numbers.} 
\sa{The definitions of the considered integration regions are detailed in sections~\ref{sec:TDCP_Expression} and~\ref{sec:application}.} 
In this article we adopt the convention \mbox{$\CP \ket{\Bz} = + \ket{\Bzb}$}. \es{This convention is equivalent to \mbox{$\C \ket{\Bz} = - \ket{\Bzb}$}.}
The mass eigenstates are defined as \mbox{$\ket{B_{1/2}} = p\ket{\Bz} \pm q\ket{\Bzb}$} with 
\begin{equation}
\qoverp=+\sqrt{\frac{{\rm M}_{12}^*-\frac{i}{2}\Gamma^*_{12}}{{\rm M}_{12}-\frac{i}{2}\Gamma_{12}}},	
\end{equation}
and the mass difference is taken such that 
\begin{equation}
\dm={\rm M}_2-{\rm M}_1 = -2\Real{\qoverp({\rm M}_{12}-i\frac{\Gamma_{12}}{2})}.	
\end{equation}

From the expression for \SKsrho in \Eqref{eq:v3-2_v2a}, the numerator is zero, and no mixing-induced \CP violation is expected, unless the \Bz and \Bzb can both decay into final states with the same photon polarisation $\lambda$. 
If the interference is non-zero, mixing-induced \CP violation may have observable effects.
\sas{The dependence of the \CP parameters from the measurement of \SKsrho on the \CP signs inherited by the decay of the kaonic resonances is studied in \secref{sec:amplitudes}.} 

As discussed in \secref{sec:introduction}, other intermediate states than \rhoKs are expected in \mbox{$\Bz \to \Kres \g \to \pip \pim \KS \g$} decays, and these need to be carefully separated.
Including all contributions, the time-dependent \CP asymmetry expression becomes 
\begin{equation}
\frac{\overline{\Gamma}_{\pip \pim \KS \g}(t) - \Gamma_{\pip \pim \KS \g}(t)}{\overline{\Gamma}_{\pip \pim \KS \g}(t) + \Gamma_{\pip \pim \KS \g}(t)} \equiv 
\SKspipi \sin(\dm t) - \CKspipi \cos(\dm t),
\label{eq:2-1}
\end{equation}
with 
\begin{eqnarray}
	\SKspipi &=&  \frac{2\Imag{\mathlarger{\qoverp} \sa{\mathlarger{\int}}\sumLR{ \AmpTotConj \AmpTotBar} \sa{dp}}}{\sa{\mathlarger{\int}} \sumLR{ \ModSq{\AmpTotBar} + \ModSq{\AmpTot} } \sa{dp}}, \label{eq:v3-2a} \\ [0.5em]
	\CKspipi &=& -\frac{\sa{\mathlarger{\int}}\sumLR{ \ModSq{\AmpTotBar} - \ModSq{\AmpTot} } \sa{dp}}{\sa{\mathlarger{\int}}\sumLR{ \ModSq{\AmpTotBar} + \ModSq{\AmpTot} } \sa{dp}}.\label{eq:v3-2b}
\end{eqnarray}
%
The \Bz  and \Bzb decay amplitudes, \AmpTot and \AmpTotBar, respectively, are now sums over the three considered intermediate states 
\begin{eqnarray}
	\AmpTot    &=& \AmpTotRhoKs    + \AmpTotKstPi    + \AmpTotKappaPi    \label{eqv1:1},\\
	\AmpTotBar &=& \AmpTotBarRhoKs + \AmpTotBarKstPi + \AmpTotBarKappaPi \label{eqv1:2}.
\end{eqnarray}
%

%% file: amplitudeRelations.tex
\section{Relations between amplitudes}\label{sec:amplitudes}

In this section, we derive the \CP sign, establishing relations among the four amplitudes \AmpTotL, \AmpTotR, \AmpTotBarL and \AmpTotBarR. 
\sas{For each resonance \Kres,} the decay amplitude \AmpTot (\AmpTotBar) can be written as the sum of products $\AmpTot=\sum_i\AmpWeak\times\AmpStrongGeneric$  ($\AmpTotBar=\sum_i\AmpWeakBar\times\AmpStrongBarGeneric$), where
\AmpWeak (\AmpWeakBar) is the decay amplitude of \Bz (\Bzb) to $\Kres\g$ ($\Kresbar\g$), and 
\AmpStrongGeneric (\AmpStrongBarGeneric) is  the decay amplitude of \Kres (\Kresbar) to the intermediate state $i$:
\begin{eqnarray}
	& \AmpTot 
		 = \AmpTotRhoKs + \AmpTotKstPi + \AmpTotKappaPi 
         =  \AmpWeak \times \left(\AmpStrongRhoKs + \AmpStrongKstPi + \AmpStrongKappaPi \right), \label{eq:v3-12}\\
	& \AmpTotBar 
		 = \AmpTotBarRhoKs + \AmpTotBarKstPi + \AmpTotBarKappaPi 
         =  \AmpWeakBar \times \left(\AmpStrongBarRhoKs + \AmpStrongBarKstPi + \AmpStrongBarKappaPi \right).\label{eq:v3-13}
\end{eqnarray}
\es{We keep $\lambda$ in the strong decay amplitude, \AmpStrongGeneric, though, as discussed later in this section, the squared amplitude does not depend on $\lambda$.}

\subsection{\boldmath{$\B \to \Kres \g$ amplitudes}}\label{sec:amplitudes_B}

First, we consider the \B decay part. 
In the SM, the $B (\Bb) \to \Kres (\Kresbar) \g$ transition comes from the penguin diagram with a top quark and a $W$ boson in the loop.
These interactions can be written by the matrix elements
\begin{eqnarray}
	&\AmpWeakRBar = \braket{\Kresbar \gR}{\Hm   }{\Bb}, \quad
 	 \AmpWeakLBar = \braket{\Kresbar \gL}{\Hp   }{\Bb}, \\
	&\AmpWeakR    = \braket{\Kres    \gR}{\Hpdag}{\B }, \quad
	 \AmpWeakL    = \braket{\Kres    \gL}{\Hmdag}{\B },  
\end{eqnarray}
where the effective Hamiltonians, at leading order in QCD, are 
\begin{eqnarray}
	&&\Hp = -\frac{G_F}{\sqrt{2}}\frac{e}{16\pi^2}\Vtb\Vtsstar \mb\Wc \left[\squarkbar \sigma^{\mu\nu}(1+ \g_5)\bquark F_{\mu\nu}\right], \\ 
	&&\Hm = -\frac{G_F}{\sqrt{2}}\frac{e}{16\pi^2}\Vtb\Vtsstar \mb\cprime \left[\squarkbar \sigma^{\mu\nu}(1- \g_5)\bquark F_{\mu\nu}\right],
\end{eqnarray}
and 
\begin{equation}
	\Wc = -\frac{1}{2}{F_2(m_t)} , \quad \cprime = -\frac{1}{2}{F_2(m_t)}\ \frac{\ms}{\mb},
\end{equation}
where $F_2$ is the Inami-Lim function that includes the top quark loop contribution~\cite{Inami:1980fz}. 
The  $\cprime$ contribution, which is proportional to a small factor $\ms/\mb$, is often neglected in the literature.\footnote{Apart from the term that is proportional to \ms, the right handed contribution, \cprime, also receives some small contributions from the charm quark loop (see Refs.~\cite{Grinstein:2004uu,Grinstein:2005nu,Khodjamirian:1997tg,Ball:2006cva,Khodjamirian:2010vf,Matsumori:2005ax} for more details).}
Including the one-loop QCD correction to this contribution, $c$ becomes simply the Wilson coefficient $C_{7\gamma}^{(0){\rm eff}}$. 
By including the right-handed contributions from new physics, the \cprime coefficient can be affected as mentioned in the introduction. 
Note that for \btosgbar transitions, the Hamiltonians are given by the Hermitian conjugate
\begin{eqnarray}
	&&\Hpdag = -\frac{G_F}{\sqrt{2}}\frac{e}{16\pi^2}\Vtbstar\Vts \mb\cstar \left[\bquarkbar \sigma^{\mu\nu}(1- \g_5)\squark F_{\mu\nu}\right], \\ 
	&&\Hmdag =-\frac{G_F}{\sqrt{2}}\frac{e}{16\pi^2}\Vtbstar\Vts  \mb\cprimestar \left[\bquarkbar \sigma^{\mu\nu}(1+ \g_5)\squark F_{\mu\nu}\right].
\end{eqnarray}
%

We can also find the relations among the \B decay amplitudes \AmpWeakLR and \AmpWeakLRBar by applying the parity (\P) and charge-conjugation (\C) operators. 
We first consider the case where \Kres is a $J^P=1^+$ state. 
Inserting the unit matrix $\Pdag\P$ yields
\begin{eqnarray}
	\AmpWeakRBar 
		&=& \braket{\Kresbar \gR}{\Pdag\P \Hm \Pdag\P}{\Bb}\nonumber \\
		&=& \eta^{\P}(B)\eta^{\P}(\Kres )\eta^{\P}(\g )(-1)^{j_{0}-j_{1}-j_{2}}\left(\cprimeoverc \right)
			\braket{\Kresbar \gL}{\Hp}{\Bb} \nonumber \\
		&=& + \left(\cprimeoverc \right)\AmpWeakLBar,
\end{eqnarray}
where $j_i$ is the total spin of the initial ($i=0$) and the final ($i=1,2 $) particles, \sa{and $\eta^{\P}(X)$ is the intrinsic parity of particle $X$}.
Here we used \mbox{$\P  \squarkbar \sigma^{\mu\nu} (1-\g_5) \bquark F_{\mu\nu} \Pdag = + \squarkbar \sigma^{\mu\nu} (1+\g_5) \bquark F_{\mu\nu}$}. 
Similarly  the relation between \AmpWeakLR and \AmpWeakLRBar can be obtained by applying a \C transformation. 
Inserting the unit matrix $\Cdag\C$ yields
\begin{eqnarray}
	\AmpWeakRBar 
	 &=& \braket{\Kresbar \gR}{\Cdag\C \Hm \Cdag\C}{\Bb} \nonumber \\
     &=& \eta^{\C}(\B)\eta^{\C}(\Kres)\eta^{\C}(\g) \left(\cprimeovercstar \right)
     	 \braket{\Kres \gR}{-\Hpdag}{\B} \nonumber \\
     &=& +\left(\cprimeovercstar \right)\AmpWeakR,
\end{eqnarray}
where \sa{$\eta^{\C}(X)$ is the charge-conjugation eigenvalue of particle $X$}. 
Here we used
  \mbox{$\C \squarkbar \sigma_{\mu\nu} (1-\g_5) \bquark F^{\mu\nu}  \Cdag = -\bquarkbar \sigma_{\mu\nu} (1-\g_5) \squark F^{\mu\nu}$}.
The phase convention of the \C transformation of \Kres is chosen to be $\C \ket{\Kres} = - \ket{\Kresbar}$ throughout this article.
\es{For the amplitudes of \B mesons decaying into $J^P=1^+$ kaonic states we finally obtain:}
\begin{equation}
\begin{minipage}{10cm}
\begin{eqnarray}
	&\AmpWeakRBar =+\left(\cprimeoverc\right)\AmpWeakLBar, \quad 
	 \AmpWeakR    =+\left(\cstarovercprimestar\right)\AmpWeakL, \quad 
	\es{\AmpWeakRBar =+\left(\CprimeoverCprimestar\right)\AmpWeakL,} &  \nonumber \\
	&\AmpWeakRBar =+\left(\cprimeovercstar\right)\AmpWeakR, \quad 
	 \AmpWeakLBar =+\left(\covercprimestar\right)\AmpWeakL.& \nonumber 
\end{eqnarray}
\end{minipage}
\label{eq:2-16}
\end{equation}
\es{Following the same formalism, for the $J^P=1^-$ and $J^P=2^+$ kaonic states we obtain:}
\begin{equation}
\begin{minipage}{10cm}
\begin{eqnarray}
	&\AmpWeakRBar = -\left(\cprimeoverc \right)         \AmpWeakLBar, \quad 
 	 \AmpWeakR    = -\left(\cstarovercprimestar \right) \AmpWeakL, \quad 
	\es{\AmpWeakRBar =-\left(\CprimeoverCprimestar\right)\AmpWeakL,} &  \nonumber \\
	&\AmpWeakRBar = +\left(\cprimeovercstar \right)     \AmpWeakR, \quad 
 	 \AmpWeakLBar = +\left(\covercprimestar \right)     \AmpWeakL. & \nonumber 
\end{eqnarray}
\end{minipage}
\label{eq:2-19}
\end{equation}
%

\subsection{\boldmath{$\Kres \to \pip \pim \KS$ amplitudes}}\label{sec:amplitudes_Kres}

First, we find a relation between \AmpStrongGeneric and \AmpStrongBarGeneric, which are related by the \C transformation.
The amplitude \AmpStrongBarGeneric corresponds to the same point in phase space as \AmpStrongGeneric, since we are interested in the interference between the two. 
To start, the amplitude of the decay $\Kres \to \rhoKs$ can be written in terms of the product of the matrix elements: 
\begin{eqnarray} 
	\AmpStrongRhoKs  &=& 
		\braket{\pip(p_1)\pim(p_2)}{\Hsprime}{\rhoz}
		\braket{\rhoKs(p_3)}{\Hs}{\Kres},  \\
	\AmpStrongBarRhoKs &=& 
		\braket{\pip(p_1)\pim(p_2)}{\Hsprime}{\rhoz}  
		\braket{\rhoKs(p_3)}{\Hs}{\Kresbar},   
\end{eqnarray}
where \Hs and \Hsprime are the Hamiltonians describing the two corresponding strong decays. 
Applying a \C transformation gives
\begin{eqnarray}
	\AmpStrongRhoKs 
		&=& \phantom{-} \braket{\pip(p_1)\pim(p_2)}{\Cdag\C \Hsprime \Cdag\C}{\rhoz} 
						\braket{\rhoz \KS(p_3)}{\Cdag\C \Hs \Cdag\C}{\Kres} \nonumber \\
		&=& \phantom{-} \braket{\pim(p_1) \pip(p_2)}{\Hsprime}{\rhoz}  
						\braket{\rhoz \KS(p_3)}{\Hs}{\Kresbar}  \nonumber  \\
		&=& - \braket{\pim(p_2) \pip(p_1)}{\Hsprime}{\rhoz}
		      \braket{\rhoz \KS(p_3)}{\Hs}{\Kresbar}, \label{eq:As_rhoKs_dev}
\end{eqnarray}
where \Hs and \Hsprime are invariant under charge conjugation. 
In this development we used $\C\ket{\rhoz}=-\ket{\rhoz}$, $\C\ket{\KS}=-\ket{\KS}$, related to the approximation $\CP\ket{\KS}=\ket{\KS}$, and $\C \ket{\Kres} = - \ket{\Kresbar}$, according to the convention given above. 
In the second line of \Eqref{eq:As_rhoKs_dev}, the \C transformation swaps the \pip and \pim momenta. 
Since $\rhoz \to \pip \pim$ is a $P$-wave decay, interchanging $p_1$ and $p_2$ leads to an overall minus sign in the third line of \Eqref{eq:As_rhoKs_dev}.
Thus, writing explicitly the momentum assignment of $\pip \pim \KS$ we obtain 
\begin{equation}
\label{eq:AmpRhoRelations}
\begin{minipage}{10cm}
\begin{eqnarray}
	\AmpStrongRhoKs (\MomNormal) 
		&=& \phantom{-}  \AmpStrongBarRhoKs(\MomSwapped) \nonumber \\
		&=&            - \AmpStrongBarRhoKs(\MomNormal). \nonumber
\end{eqnarray}
\end{minipage}
\end{equation} 
The amplitude describing the $\Kres \to \Kst\pi$ process can be written as
\begin{eqnarray} 
	\AmpStrongKstPi 
		&=& \braket{\KS(p_3)\pip(p_1)}{\Hsprime}{\Kstp} 
			\braket{\KstPi(p_2)}{\Hs}{\Kres}, \nonumber  \\
	\AmpStrongBarKstPi 
		&=& \braket{\KS(p_3)\pim(p_2)}{\Hsprime}{\Kstm} 
			\braket{\KstPibar(p_1)}{\Hs}{\Kresbar},
\end{eqnarray}
and applying the \C transformation results in 
\begin{eqnarray}
	\AmpStrongKstPi(\MomNormal) 
		&=& \braket{\KS(p_3) \pip(p_1)}{\Cdag\C \Hsprime \Cdag\C}{\Kstp} 
			\braket{\Kst \pim(p_2)}{\Cdag\C \Hs \Cdag\C}{\Kres} \nonumber \\
 		&=& \braket{\KS(p_3) \pim(p_1)}{\Hsprime}{\Kstm}
 			\braket{\Kstm \pip(p_2)}{\Hs}{\Kresbar} \nonumber \\
		&=& \AmpStrongBarKstPi(\MomSwapped), \label{eq:AmpKstRelations} 
\end{eqnarray}
where the momentum assignment is explicitly written for clarity.  
This reflects the general strong-interaction dynamics, where no simple relation allows to interchange $p_1$ and $p_2$ when
the \sas{\pip} and \pim are swapped.
Thus, the relation between $\AmpStrongKstPi(\MomNormal)$ and $\AmpStrongBarKstPi(\MomNormal)$ is unknown, except for the trivial case where $p_1=p_2$. 
A similar conclusion applies when considering the process $\Kres \to \swave\pi$.

It is possible to obtain relations between \AmpStrongLGeneric and \AmpStrongRGeneric and between \AmpStrongLBarGeneric and \AmpStrongRBarGeneric. 
Indeed, since the decay of the kaonic resonances only depend on the strong interaction, the former and the latter relations are expected to be the same. 
Furthermore, since decays with left- and right-handed photons do not interfere, only relations between products of amplitudes with the same photon polarisation are needed.
The explicit computation for three kaonic resonances shows that, after integrating over the decay angles, these products do not depend on the left or right polarisation of the resonances:
\begin{equation}
\begin{minipage}{10cm}
\[
\AmpStrongRGenericConj \AmpStrongR^j       = \AmpStrongLGenericConj \AmpStrongL^j , \quad 
\AmpStrongRBarGenericConj \AmpStrongRBar^j = \AmpStrongLBarGenericConj \AmpStrongLBar^j,
\]
\end{minipage}
\label{eq:v3-30}
\end{equation}
where $i, j=\rhoKs, \Kstpm \pi^\mp, \swave^\pm\pi^\mp$. 

%% file: TDCP_Expression.tex
\section{\boldmath{Expression of the time-dependent \CP asymmetry}}
\label{sec:TDCP_Expression}

The expressions of the mixing-induced \CP violation parameters, \SKsrho and \SKspipi, given in \Eqsref{eq:v3-2_v2a}{eq:v3-2a}, respectively, can be rewritten using the relations between the amplitudes describing \mbox{$\Bz \to \Kres \g \to \pip \pim \KS \g$} decays, obtained in \secref{sec:amplitudes}. 
The aim is to express \SKsrho and \SKspipi in terms of amplitudes corresponding to a single $B$-flavour (choosing \Bz) and a single polarisation (choosing $\lambda=L$).
The squared \Bz and \Bzb amplitudes, $\ModSq{\AmpTot}$ and $\ModSq{\AmpTotBar}$, respectively, are written \sas{for a single \Kres contribution} as
\begin{eqnarray}
\ModSq{\AmpTot} & = & 
\ModSq{\AmpTotRhoKs}+
\ModSq{\AmpTotKstPi}+
\ModSq{\AmpTotKappaPi}  \nonumber \\ 
&& +2\Real{ \AmpTotRhoKsConj \AmpTotKstPi  } 
   +2\Real{ \AmpTotRhoKsConj \AmpTotKappaPi}\,,  \nonumber\\ 
&& +2\Real{ \AmpTotKstPiConj \AmpTotKappaPi }\,,  \\
\label{eq:v3-7} \nonumber \\
\ModSq{\AmpTotBar} & = &
\ModSq{\AmpTotBarRhoKs} +
\ModSq{\AmpTotBarKstPi}+
\ModSq{\AmpTotBarKappaPi}  \nonumber \\
&& +2\Real{ \AmpTotBarRhoKsConj \AmpTotBarKstPi }
   +2\Real{ \AmpTotBarRhoKsConj \AmpTotBarKappaPi }\,,\nonumber\\ 
&& +2\Real{ \AmpTotBarKstPiConj \AmpTotBarKappaPi }\,,  
\label{eq:v3-6}
\end{eqnarray}
and the cross term by
\begin{eqnarray}
\AmpTotConj \AmpTotBar & = &
 \AmpTotRhoKsConj   \AmpTotBarRhoKs +
 \AmpTotKstPiConj   \AmpTotBarKstPi + 
 \AmpTotKappaPiConj \AmpTotBarKappaPi \nonumber \\
&& + \left[ \AmpTotRhoKsConj \AmpTotBarKstPi   + \AmpTotKstPiConj   \AmpTotBarRhoKs \right] \nonumber \\
&& + \left[ \AmpTotRhoKsConj \AmpTotBarKappaPi + \AmpTotKappaPiConj \AmpTotBarRhoKs \right] \nonumber \\
&& + \left[ \AmpTotKstPiConj   \AmpTotBarKappaPi +  \AmpTotKappaPiConj  \AmpTotBarKstPi \right], 
\label{eq:v3-5}
\end{eqnarray}
\es{where the shortened notation $\ModSq{\AmpTot}$ and $\ModSq{\AmpTotBar}$ are used instead of $\ModSq{\AmpTot(\MomNormal)}$ and $\ModSq{\AmpTotBar(\MomNormal)}$, respectively.}

The next step consists in replacing, in \Eqsref{eq:v3-6}{eq:v3-5}, the \Bzb decay amplitudes by those corresponding to the \Bz decay.  This is done by using the results obtained in \secref{sec:amplitudes} for the \CP signs of the \B decay part, in~\Eqsref{eq:2-16}{eq:2-19}, and the \Kres decay part, in~\Eqsref{eq:AmpRhoRelations}{eq:AmpKstRelations}. For $\lambda=L$ we obtain
\begin{eqnarray}
\ModSq{\AmpTotBarL(\MomNormal)} & = &
\left| \covercprimestar \right|^2\left[
   \ModSq{\AmpTotLRhoKs(\MomSwapped)}
  +\ModSq{\AmpTotLKstPi(\MomSwapped)}
  +\ModSq{\AmpTotLKappaPi(\MomSwapped)} \right. \nonumber \\
&&\phantom{\ModSq{\covercprimestar}}\;\; 
   +2\Real{\AmpTotLRhoKsConj (\MomSwapped)\AmpTotLKstPi(\MomSwapped)}\nonumber \\
&&\phantom{\ModSq{\covercprimestar}}\;\;    
+2\Real{\AmpTotLRhoKsConj(\MomSwapped) \AmpTotLKappaPi(\MomSwapped)}  \nonumber \\
&&\phantom{\ModSq{\covercprimestar}}\;\; \left.   
+2\Real{\AmpTotLKstPi(\MomSwapped) \AmpTotLKappaPi(\MomSwapped)} \right] \nonumber \\
&=&\ModSq{\covercprimestar} \ModSq{\AmpTotL(\MomSwapped)}\,,
\label{eq:v4-25}
\end{eqnarray}
and
\begin{eqnarray}
 \AmpTotLConj(\MomNormal) \AmpTotBarL(\MomNormal) &=& \left(\covercprimestar\right) \AmpTotLConj(\MomNormal) \AmpTotL(\MomSwapped) .
\end{eqnarray}
Furthermore, we express all the amplitudes in term of one polarisation, choosing $\lambda=L$.
To do so, we obtain relations between left and right amplitudes using~\Eqref{eq:v3-30}, together with~\Eqsref{eq:2-16}{eq:2-19}
\begin{eqnarray}
\AmpTotRConj(\MomNormal) \AmpTotBarR(\MomNormal) & = & \AmpTotLConj(\MomNormal) \AmpTotBarL(\MomNormal) \nonumber  \\
\ModSq{\AmpTotR(\MomNormal)}    &=& \ModSq{\covercprime}\ModSq{\AmpTotL(\MomNormal)} \nonumber \\
\ModSq{\AmpTotBarR(\MomNormal)} &=& \ModSq{\frac{c^\prime}{c}}  \ModSq{\AmpTotBarL(\MomNormal)}. 
\label{eq:4-3}
\end{eqnarray}
Using these relations, \Eqref{eq:v3-2a} can be re-written as
\begin{eqnarray}
\SKspipi &=& 4{\rm Im}\left(\frac{q}{p}\frac{\Wc \cprime }{\ModSq{\Wc}+\ModSq{\cprime}} 
\vphantom{\frac{\mathlarger{\int}}{\mathlarger{\int}}} \right. 
\qquad\qquad\qquad\qquad\qquad\qquad\qquad\qquad\qquad\qquad\qquad\qquad\quad\quad\quad \nonumber \\
&& \left. \times \frac{\mathlarger{\int}  {\sum_{i,j} \left[\AmpTotLIConj(\MomNormal)\AmpTotLJ(\MomSwapped) \right]} dp }{\mathlarger{\int} {\sum_{i,j} \left[\AmpTotLIConj(\MomNormal)\AmpTotLJ(\MomNormal) + \AmpTotLIConj(\MomSwapped)\AmpTotLJ(\MomSwapped) \right]} dp } \right) \nonumber \\ 
 &=& 4{\rm Im}\left(\frac{q}{p}\frac{\Wc \cprime }{\ModSq{\Wc}+\ModSq{\cprime}} 
 \vphantom{\frac{\mathlarger{\int}}{\mathlarger{\int}}} \right. 
 \qquad\qquad\qquad\qquad\qquad\qquad\qquad\qquad\qquad\qquad\qquad\qquad\quad\quad\quad \nonumber \\
&& \left. \times \frac{\mathlarger{\int}  {\sum_{i,j} \left[\AmpStrongGIConj(\MomNormal)\AmpStrongGJ(\MomSwapped) \right]} dp }{\mathlarger{\int}  {\sum_{i,j} \left[\AmpStrongGIConj(\MomNormal)\AmpStrongGJ(\MomNormal) + \AmpStrongGIConj(\MomSwapped)\AmpStrongGJ(\MomSwapped) \right]} dp } \right), 
\label{eq:sec-4-SKspipi}
\end{eqnarray}
where $i,j$ run over ${\rhoKs, \KstPi, \KappaPi}$. Here we also used the fact that the weak-decay part of the total amplitude can be factored out, as $\AmpTotLIConj(\MomNormal) = \AW_L \AmpStrongGIConj(\MomNormal)$, and thus cancels out in the ratio.
The notation $\AmpStrongGIConj(\MomNormal)$ corresponds to the strong part of the amplitude, averaged over the $K_{\rm res}$ helicity states.
Similarly, \Eqref{eq:v3-2b} can be re-written as
\begin{equation}
\label{eq:sec-4-CKspipi}
\CKspipi =  \frac{\mathlarger{\int} {\sum_{i}\left[\ModSq{\AmpStrongGI(\MomNormal)} - \ModSq{\AmpStrongGI(\MomSwapped)}\right]} dp }{\mathlarger{\int}{\sum_{i}\left[\ModSq{\AmpStrongGI(\MomNormal)}\right] +  \ModSq{\AmpStrongGI(\MomSwapped)} } dp}, 
\end{equation}
where the phase-space $dp$ represents the two Dalitz-plot variables and the $K\pi\pi$ invariant mass. 

We emphasise that the expressions of \SKspipi and \CKspipi obtained in \Eqsref{eq:sec-4-SKspipi}{eq:sec-4-CKspipi}, respectively, are independent of the intermediate kaonic resonance. 
Furthermore, the interferences of $J^P = {1^+, 1^-, 2^+}$ kaonic resonances cancel out after integrating over the polar and the azimuthal angles of the photon direction with respect to the $K\pi\pi$ decay plane (i.e. $\theta$ and $\phi$ in~\cite{Kou:2016iau}). 
This means that \Eqref{eq:sec-4-SKspipi} can be extended to yield the value of \SKspipi including the contributions from all kaonic resonances by simply replacing the numerator and the denominator of the last line of \Eqref{eq:sec-4-SKspipi} by the sum of them for $J^P = {1^+, 1^-, 2^+}$. 
Thus, \Eqref{eq:sec-4-SKspipi} can be used by experimental studies in two ways: i) if the different kaonic resonances can be distinguished experimentally, the amplitudes in \Eqref{eq:sec-4-SKspipi} can be considered as those of a given kaonic resonance decaying into the corresponding isobars; ii) if the kaonic resonances are not distinguished experimentally, then the amplitudes can be considered as sums over all the kaonic resonances decaying into the corresponding isobars. 

Expanding the sum over the hadronic amplitudes in the expression of \SKspipi in \Eqref{eq:sec-4-SKspipi}, we obtain
\begin{eqnarray}
 \sum_{i,j} \left[\AmpStrongGIConj(\MomNormal)\AmpStrongGJ(\MomSwapped) \right]  &=& 
  \AmpStrongGRhoKsConj(\MomNormal) \AmpStrongGRhoKs(\MomSwapped) \label{eq:v4-24} \\ [-1.0em]
&& + \AmpStrongGKstPiConj(\MomNormal) \AmpStrongGKstPi(\MomSwapped) \nonumber \\
&& + \AmpStrongGKappaPiConj(\MomNormal) \AmpStrongGKappaPi(\MomSwapped) \nonumber \\
&& + \AmpStrongGRhoKsConj(\MomNormal) \AmpStrongGKstPi(\MomSwapped) \nonumber \\
&& + \AmpStrongGKstPiConj(\MomNormal) \AmpStrongGRhoKs(\MomSwapped) \nonumber \\
&& + \AmpStrongGRhoKsConj(\MomNormal) \AmpStrongGKappaPi(\MomSwapped) \nonumber \\
&& + \AmpStrongGKappaPiConj(\MomNormal) \AmpStrongGRhoKs(\MomSwapped) \nonumber \\
&& + \AmpStrongGKstPiConj (\MomNormal)  \AmpStrongGKappaPi(\MomSwapped) \nonumber \\  
&& + \AmpStrongGKappaPiConj(\MomNormal)  \AmpStrongGKstPi(\MomSwapped)  \nonumber \\ [0.3em]
&=& - \ModSq{\AmpStrongGRhoKs(\MomNormal)}    \label{eq:v4-24-2}   \\
&&  + {\AmpStrongGKstPiConj(\MomNormal) \AmpStrongGKstPi(\MomSwapped)}   \nonumber \\
&&  + {\AmpStrongGKappaPiConj(\MomNormal) \AmpStrongGKappaPi(\MomSwapped)} \nonumber \\
&&  -2\Real{\AmpStrongGRhoKsConj(\MomNormal) \AmpStrongGKstPi(\MomNormal)} \nonumber \\
&&  -2\Real{\AmpStrongGRhoKsConj(\MomNormal) \AmpStrongGKappaPi(\MomNormal)}   \nonumber \\
&&  +\AmpStrongGKstPiConj (\MomNormal)  \AmpStrongGKappaPi(\MomSwapped) \nonumber \\  
&&  +\AmpStrongGKappaPiConj(\MomNormal)  \AmpStrongGKstPi(\MomSwapped) , \nonumber 
\end{eqnarray}
and
\begin{eqnarray}
 \sum_{i,j} \left[\AmpStrongGIConj(\MomNormal)\AmpStrongGJ(\MomNormal) \right]  &=& 
  \ModSq{\AmpStrongGRhoKs(\MomNormal)}  \label{eq:v4-24-3} \\ [-1.0em]
&&  + \ModSq{\AmpStrongGKstPi(\MomNormal)} \nonumber \\
&&  + \ModSq{\AmpStrongGKappaPi(\MomNormal)} \nonumber \\
&&  +2\Real{\AmpStrongGRhoKsConj(\MomNormal) \AmpStrongGKstPi(\MomNormal)} \nonumber \\
&&  +2\Real{\AmpStrongGRhoKsConj(\MomNormal) \AmpStrongGKappaPi(\MomNormal)}   \nonumber \\
&&  +2\Real{\AmpStrongGKstPiConj(\MomNormal) \AmpStrongGKappaPi(\MomNormal)} .  \nonumber 
\end{eqnarray}
The minus signs in \Eqref{eq:v4-24-2} originate from the relation in \Eqref{eq:AmpRhoRelations}.
On the other hand, as shown in \secref{sec:amplitudes}, there is no general symmetry relation between $\AmpStrongGKstPi(\MomNormal)$ and $\AmpStrongGKstPi(\MomSwapped)$, and similarly for \swave. 
Thus the second, third, sixth and seventh terms in \Eqref{eq:v4-24-2} cannot be further simplified. 

From \Eqsref{eq:v4-24-2}{eq:v4-24-3}, it follows that in a time-dependent amplitude analysis, the \CP asymmetry measurement can be directly related to the photon polarisation for the \rhoKs amplitude as
\begin{equation}
\SKsrho = -\frac{2\Imag{\qoverp \Wc \cprime}}{\ModSq{\Wc}+\ModSq{\cprime}},
\label{eq:SrhoKs_reduced}
\end{equation} 
\sa{which means that the time-dependent \CP asymmetry of \nChannelRhoz decays has an opposite sign of that of $\Bz \to \piz \KS \gamma$ decays (see Appendix~\ref{App:CompOtherModes}),  i.e. $\SKsrho = - \mathcal{S}_{\piz\KS\gamma}$. 
Note that since both decay channels arise from the same quark-level transition, $b\to s\gamma$, and that all the hadronic \sas{effects} cancel out in this formula, this equality is valid at a high precision.}  

%% file: appliExpResults.tex
\section{\boldmath{Proposed experimental strategies}}
\label{sec:application}

In this section two methods are described to obtain information on photon polarisation from mixing-induced \CP violation parameter measurements.
The first, suitable in the context of limited-size data samples, like those used by the \babar and Belle collaborations, is described in \secref{sec:lowStatStrategy}.
\sas{Using a similar logic to that employed by \babar and Belle, this is the first time that a clear theoretical development of the \CP-violation parameters expression is proposed in this context.}
Then, in \secref{sec:highStatStrategy}, we propose a novel method that is better suited to a larger data-sample, as expected in the \superbelle experiment.

\subsection{\boldmath{Phase-space integrated analysis}}
\label{sec:lowStatStrategy}

A time-dependent amplitude analysis to extract the \CP asymmetries for individual resonances is currently not feasible at the \B factories due to the limited sizes of the available data samples. 
Indeed, only the \CP asymmetry of the full $K\pi\pi$ system is measured~\cite{Li:2008qma, Sanchez:2015pxu}.
After integration over the whole Dalitz plane of the $K\pi\pi$ system, the expressions of the time-dependent \CP asymmetry parameters can be expressed as
\begin{eqnarray}
\label{eq:sec-5-SKspipi}
\SKspipi &=& \frac{2\Imag{\qoverp \Wc \cprime }}{\ModSq{\Wc}+\ModSq{\cprime}}  \frac{\mathlarger{\int}_{\rm tot} {\Real{ \AmpStrongGConj(\MomNormal)\AmpStrongG(\MomSwapped) }} dp }{\mathlarger{\int}_{\rm tot} { \ModSq{ \AmpStrongG(\MomNormal)}} dp }, \\
\CKspipi &=& 0.
\end{eqnarray}
As shown in \figref{fig:KstReIm} and \figref{fig:AllReIm}, the imaginary part of $\AmpStrongGConj(\MomNormal)\AmpStrongG(\MomSwapped)$ cancels after integration over the whole Dalitz plane and thus does not appear in the numerator of \Eqref{eq:sec-5-SKspipi}.
Furthermore, the factor 2 in the numerator of \Eqref{eq:sec-5-SKspipi} originates from the fact that \mbox{$\int_{\rm tot}\ModSq{ \AmpStrongG(\MomNormal)}  dp = \int_{\rm tot}\ModSq{ \AmpStrongG(\MomSwapped)} dp$}.
Then, the mixing-induced \CP asymmetry of the \rhoKs mode, given in \Eqref{eq:SrhoKs_reduced}, is obtained via the dilution factor
\begin{eqnarray}
\D & \equiv & \frac{\SKspipi}{\SKsrho} \label{eq:sec-5-Dil}\\
  &=& -  \frac{\mathlarger{\int}_{\rm tot} {\Real{ \AmpStrongGConj(\MomNormal)\AmpStrongG(\MomSwapped) }} dp }{\mathlarger{\int}_{\rm tot} { \ModSq{ \AmpStrongG(\MomNormal)}} dp }, \nonumber
\end{eqnarray}
where 
\begin{eqnarray}
 \Real{ \AmpStrongGConj(\MomNormal)\AmpStrongG(\MomSwapped) } &=& - \ModSq{\AmpStrongGRhoKs(\MomNormal)}    \label{eq:sec-5-RealMMbar}   \\
&&  -2\Real{\AmpStrongGRhoKsConj(\MomNormal) \AmpStrongGKstPi(\MomNormal)} \nonumber \\
&&  -2\Real{\AmpStrongGRhoKsConj(\MomNormal) \AmpStrongGKappaPi(\MomNormal)}   \nonumber \\
&&  + \Real{\AmpStrongGKstPiConj(\MomNormal) \AmpStrongGKstPi(\MomSwapped)}   \nonumber \\
&&  + \Real{\AmpStrongGKappaPiConj(\MomNormal) \AmpStrongGKappaPi(\MomSwapped)} \nonumber \\
&&  +\Real{\AmpStrongGKstPiConj (\MomNormal)  \AmpStrongGKappaPi(\MomSwapped)} \nonumber \\  
&&  +\Real{\AmpStrongGKappaPiConj(\MomNormal)  \AmpStrongGKstPi(\MomSwapped)} , \nonumber 
\end{eqnarray}
and
\begin{eqnarray}
 \ModSq{ \AmpStrongG(\MomNormal)} &=& 
  \ModSq{\AmpStrongGRhoKs(\MomNormal)}  \label{eq:sec-5-MSq} \\
&&  + \ModSq{\AmpStrongGKstPi(\MomNormal)} \nonumber \\
&&  + \ModSq{\AmpStrongGKappaPi(\MomNormal)} \nonumber \\
&&  +2\Real{\AmpStrongGRhoKsConj(\MomNormal) \AmpStrongGKstPi(\MomNormal)} \nonumber \\
&&  +2\Real{\AmpStrongGRhoKsConj(\MomNormal) \AmpStrongGKappaPi(\MomNormal)}.   \nonumber
\end{eqnarray}
Note that the last term in \Eqref{eq:v4-24-3}, which represents the interference between the \KstPi and \KappaPi amplitudes, does not appear in \Eqref{eq:sec-5-MSq}. 
This is due to the fact that this term cancels after integration over the \sa{Dalitz plane}, as the interference between $P$- and $S$-waves \sa{is a linear combination of odd- and even-order Legendre polynomials}. 

The measurement of the dilution factor of Eq.~\eqref{eq:sec-5-Dil} can be performed via a time-integrated analysis.
As proposed in Refs.~\cite{Li:2008qma, Sanchez:2015pxu}, in order to obtain the best sensitivity for \D, its value can be measured from an amplitude analysis of $\Bp \to \Kresp \g \to \Kp \pim \pip \g$ decays,\footnote{\es{Charge conjugation is implicit here}.} assuming isospin symmetry.
Indeed, a larger data sample is expected for the final state $\Kp \pim \pip \g$  comparing to the neutral isospin partner $\KS \pim \pip \g$, due to a larger branching fraction, as well as better experimental reconstruction and selection efficiencies. 

\sas{From the detailed expressions in Eq. 46 and 47, which are explicitly given for the first time in this paper, it is clear that} the expression of \SKspipi, given in \Eqref{eq:sec-4-SKspipi}, and hence of the dilution factor, are valid across the whole $K\pi\pi$ phase space and can be integrated. 
\es{On the other hand, as the sensitivity to the \CP parameters is higher for larger values of the dilution factor, an optimised integration region needs to be considered. 
For instance, the $K_1(1270)$ has a larger branching fraction to $\rhoKs$, which leads to a larger dilution factor comparing to higher-spin resonances. 
Thus, an optimised set of cuts in the \mKpipi spectrum needs to be considered when measuring the dilution factor. 
Note that any requirements on the phase space must be symmetric under $\pi^+\leftrightarrow \pi^-$ interchange.}

We emphasise that the measurement of the dilution factor \D, which does not require the study of \CP asymmetries but only that of the intermediate resonance amplitudes, can be obtained independently, for instance from the LHCb experiment, benefiting from a larger data sample of $\Bp \to \Kp \pim \pip \g$ decays comparing to the \B factories.

\subsection{\boldmath{Time-dependent amplitude analysis}}
\label{sec:highStatStrategy}

Considering that a larger data sample is available, as that expected in \superbelle, 
we assume that a time-dependent amplitude analysis of \mbox{$\Bz \to \Kres \g \to \pip \pim \KS \g$} decays becomes feasible.
In this section, we show that considering different regions of the $K\pi\pi$ Dalitz plane separately, provides more information that significantly improves the sensitivity to new-physics contributions to the photon polarisation.

The expression of \SKspipi can be re-written with the integration being performed over a  region in the Dalitz plane, $\delta\dalitz$, such as
\begin{eqnarray}
\SKspipiDalitz &=&  4\Imag{\frac{q}{p}\frac{\xi }{1 + \ModSq{\xi}}\frac{\mathlarger{\int}_{\delta \dalitz} \ASDalitzConjNormal\ASDalitzSwapped dp}{\mathlarger{\int}_{\delta \dalitz} \ModSq{\ASDalitzNormal} + \ModSq{\ASDalitzSwapped}  dp}} ,  \label{eq:sec-5.2-SKspipi}
\end{eqnarray}
using, for simplicity, the conventions
\begin{eqnarray}
\ASDalitzConjNormal\ASDalitzSwapped &=& \sum_{i,j} \left[\AmpStrongGIConj(\MomNormal)\AmpStrongGJ(\MomSwapped) \right], \nonumber \\
\ModSq{\ASDalitzNormal}+ \ModSq{\ASDalitzSwapped} &=& 	{\sum_{i,j} \left[\AmpStrongGIConj(\MomNormal)\AmpStrongGJ(\MomNormal) + \AmpStrongGIConj(\MomSwapped)\AmpStrongGJ(\MomSwapped) \right]}, \nonumber \\
\frac{\xi }{1 + \ModSq{\xi}} &=&\frac{\Wc \cprime }{\ModSq{\Wc}+\ModSq{\cprime}} ,\nonumber
\end{eqnarray}
where we introduced the notation $\xi \equiv \cprime / \cstar $ as the ratio of right- to left-handed amplitudes, \sa{and where the expression of the amplitudes are taken from \Eqsref{eq:v4-24-2}{eq:v4-24-3}.}
Writing the hadronic decay amplitude as
\begin{equation}
\ASDalitzNormal = \Mod{\ASDalitzNormal}e^{i\delta^\dalitz_{\OrderNormal}} , \qquad \ASDalitzSwapped = \Mod{\ASDalitzSwapped}e^{i\delta^\dalitz_{\OrderSwapped}}  , 
\end{equation}
the real and imaginary parts of the hadronic contribution in \Eqref{eq:sec-5.2-SKspipi} can be expressed as
\begin{eqnarray}
\hspace{-10pt}
	\frac{\mathlarger{\int}_{\delta \dalitz} \ASDalitzConjNormal\ASDalitzSwapped dp}{\mathlarger{\int}_{\delta \dalitz} \ModSq{\ASDalitzNormal} + \ModSq{\ASDalitzSwapped} dp} &=& \phantom{+ i}
				\es{\underbrace{ 
							\frac{\mathlarger{\int}_{\delta \dalitz} \Mod{\ASDalitzNormal}\Mod{\ASDalitzSwapped} 
							\cos(\delta^\dalitz_{\OrderSwapped}-\delta^\dalitz_{\OrderNormal}) dp }{\mathlarger{\int}_{\delta \dalitz}\ModSq{\ASDalitzNormal} + \ModSq{\ASDalitzSwapped} dp} 
							}_\mathlarger{{\equiv {\rm a}^{\delta\dalitz}}}} \nonumber \\ [0.6em]
			&& + i \es{\underbrace{ 
							\frac{\mathlarger{\int}_{\delta \dalitz} \Mod{\ASDalitzNormal}\Mod{\ASDalitzSwapped}  \sin(\delta^\dalitz_{\OrderSwapped}-\delta^\dalitz_{\OrderNormal}) dp}{\mathlarger{\int}_{\delta \dalitz} \ModSq{\ASDalitzNormal} + \ModSq{\ASDalitzSwapped} dp}
							}_\mathlarger{{\equiv {\rm b}^{\delta\dalitz}}}} , \label{eq:sec-5.2-ReIm_Amp}
\end{eqnarray}
so that \Eqref{eq:sec-5.2-SKspipi} can be re-written as 
\begin{eqnarray}
\SKspipiDalitz &=&  4\Imag{\frac{q}{p}\frac{\xi }{1 + \ModSq{\xi}}} {\rm a}^{\delta\dalitz}  + 4\Real{\frac{q}{p}\frac{\xi }{1 + \ModSq{\xi}}}  {\rm b}^{\delta\dalitz}  \qquad\qquad\qquad\qquad\qquad \label{eq:sec-5.2-SKspipi_ReIm}  \\
			   &=& \frac{4}{1+\ModSq{\xi}}  \Big(
			   		{\rm a}^{\delta\dalitz} \left[ {\rm Im}\xi \cos 2\beta - {\rm Re}\xi \sin 2\beta  \right]
			   		+ 
			   		{\rm b}^{\delta\dalitz} \left[ {\rm Re}\xi  \cos 2\beta +  {\rm Im}\xi  \sin 2\beta\right]
			   \Big)   \nonumber \\
			   &=& \frac{4}{1+\ModSq{\xi}}  \Big(
			   		{\rm Re}\xi \left[  {\rm b}^{\delta\dalitz} \cos 2\beta -  {\rm a}^{\delta\dalitz} \sin 2\beta \right]
			   		+ 
			   		{\rm Im}\xi \left[ {\rm a}^{\delta\dalitz} \cos 2\beta  + {\rm b}^{\delta\dalitz} \sin 2\beta  \right]
			   \Big) .  \nonumber
\end{eqnarray}
Thus, measuring the real and imaginary parts of \es{the hadronic part in \Eqref{eq:sec-5.2-ReIm_Amp}} in different regions of the Dalitz plane provides a more precise determination of $\xi$. 
Indeed, in a similar way as in Ref.~\cite{Giri:2003ty}, it is possible to define symmetric regions of the Dalitz plane: $I$ above the bisector line $m_{13}-m_{23}$ and $\overline{I}$ below. 
In these symmetric regions, the relations
\begin{equation}
	{\rm a}^I  = {\rm a}^{\overline{I}} , \quad \textrm{ and } \quad {\rm b}^I = - {\rm b}^{\overline{I}} ,
\end{equation}
hold, from which the following relations are obtained:
\begin{eqnarray}
\SP &\equiv &  \SKspipiDalitzI + \SKspipiDalitzIb =  \frac{8 }{1+\ModSq{\xi}}  \left( {\rm Im}\xi \cos 2\beta - {\rm Re}\xi \sin 2\beta \right) {\rm a}^I, \qquad\qquad \label{eq:SKspipiDalitzI} \\
\SM &\equiv & \SKspipiDalitzI - \SKspipiDalitzIb =  \frac{8 }{1+\ModSq{\xi}} \left( {\rm Re}\xi  \cos 2\beta +  {\rm Im}\xi  \sin 2\beta \right) {\rm b}^I.  \label{eq:SKspipiDalitzIb}
\end{eqnarray}
From \Eqsref{eq:SKspipiDalitzI}{eq:SKspipiDalitzIb} it follows that by measuring separately the time-dependent \CP asymmetries in the regions $I$ and $\overline{I}$ it becomes possible to independently constrain the real and  imaginary parts of $\xi$. 
Note that \Eqref{eq:SKspipiDalitzI} is strictly equivalent to \Eqref{eq:sec-5-SKspipi}, and that $\D = -2 {\rm a}^I$.
Using \Eqsref{eq:SKspipiDalitzI}{eq:SKspipiDalitzIb}, ${\rm Re}\xi$ and ${\rm Im}\xi$ are expressed as
\begin{eqnarray}
\frac{{\rm Re}\xi}{1+\ModSq{\xi}} &=& \frac{1}{8} \left(\frac{ \SM }{{\rm b}^I } \cos 2\beta - \frac{ \SP }{{\rm a}^I } \sin2\beta \right), \label{eq:ReXi} \\
\frac{{\rm Im}\xi}{1+\ModSq{\xi}} &=& \frac{1}{8} \left(\frac{ \SM }{{\rm b}^I } \sin 2\beta + \frac{ \SP }{{\rm a}^I } \cos 2\beta \right). \label{eq:ImXi} 
\end{eqnarray}
\begin{figure}[t]
\begin{center}
\includegraphics[width = 0.45\linewidth]{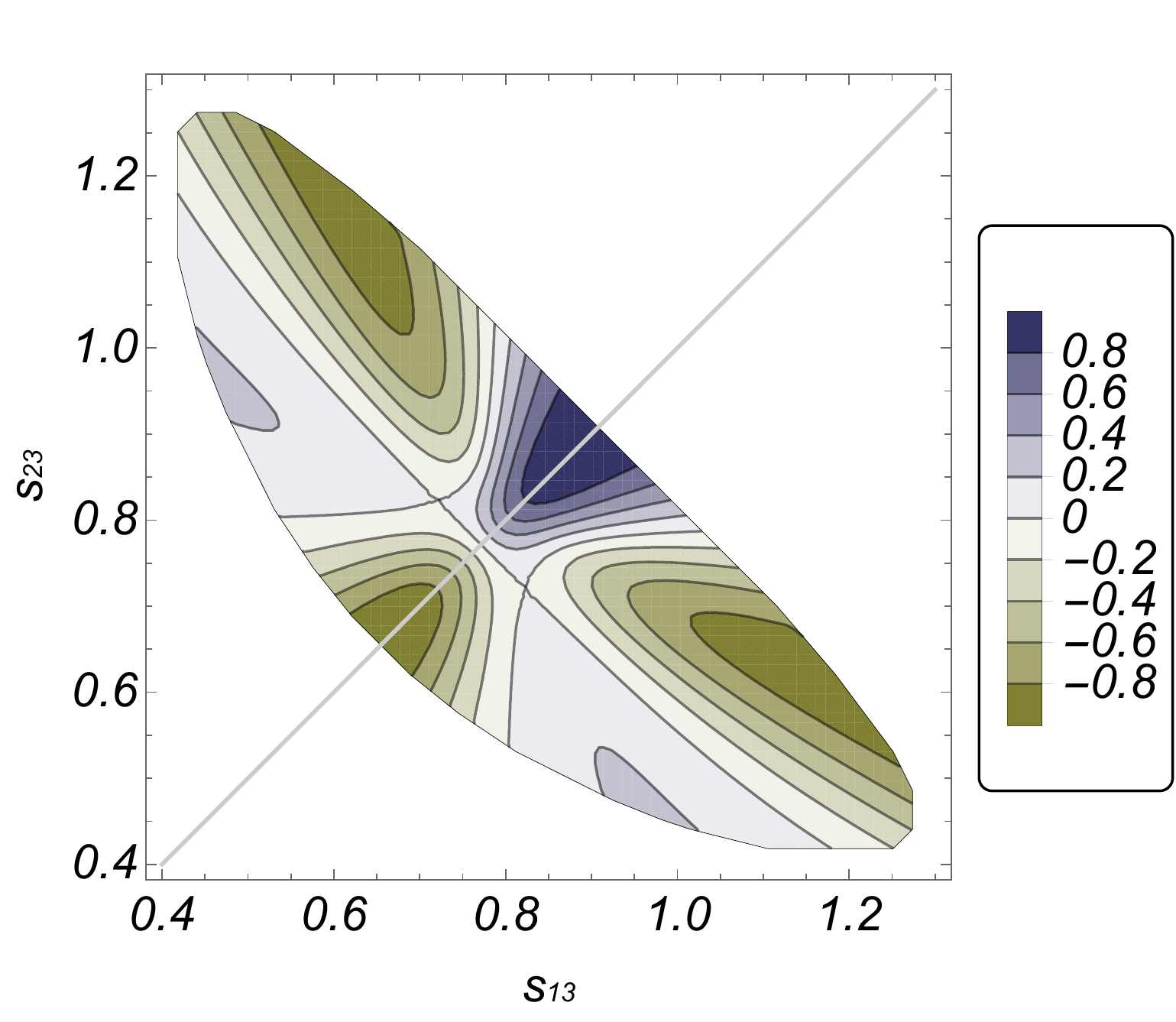}
\includegraphics[width = 0.45\linewidth]{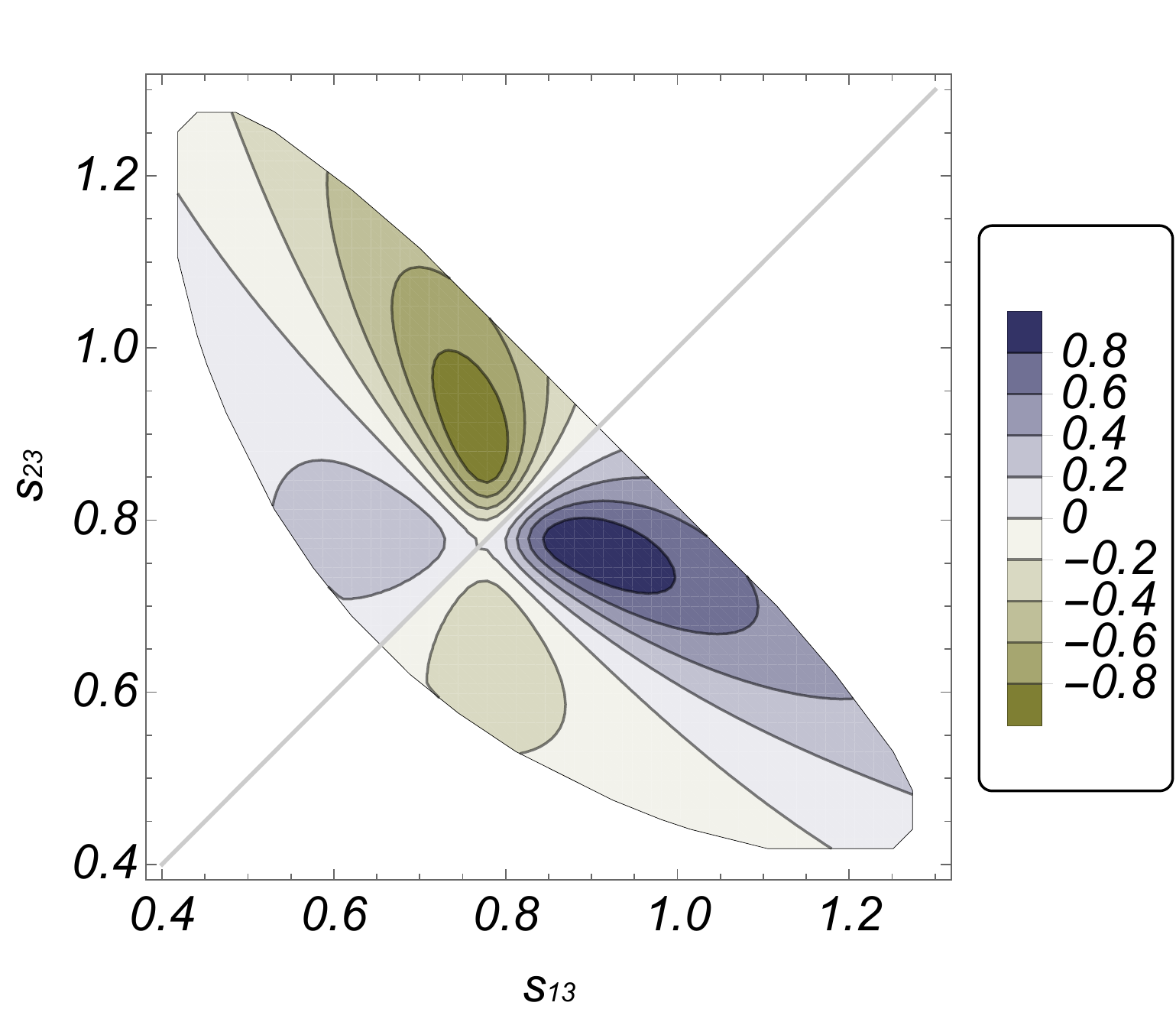}
\caption{\es{Normalised distributions of the real (left) and imaginary (right) parts of  \Eqref{eq:sec-5.2-ReIm_Amp} with  \mbox{$K_1(1270) \to \Kst\pi \to \pip\pim\KS$ decay amplitudes}. 
While the real part (${\rm a}^{\delta\dalitz}$) is symmetric with respect to the Dalitz-plane bisector, the imaginary part (${\rm b}^{\delta\dalitz}$) exhibits an anti-symmetric pattern. } 
The axes correspond to $s_{13} = m^2_{\pip\KS}$ and $s_{23} = m^2_{\pim\KS}$ in \gevcccc. A similar behaviour is observed for all kaonic resonances.}
\label{fig:KstReIm}
\end{center}
\end{figure}
\begin{figure}[t]
\begin{center}
\includegraphics[width = 0.45\linewidth]{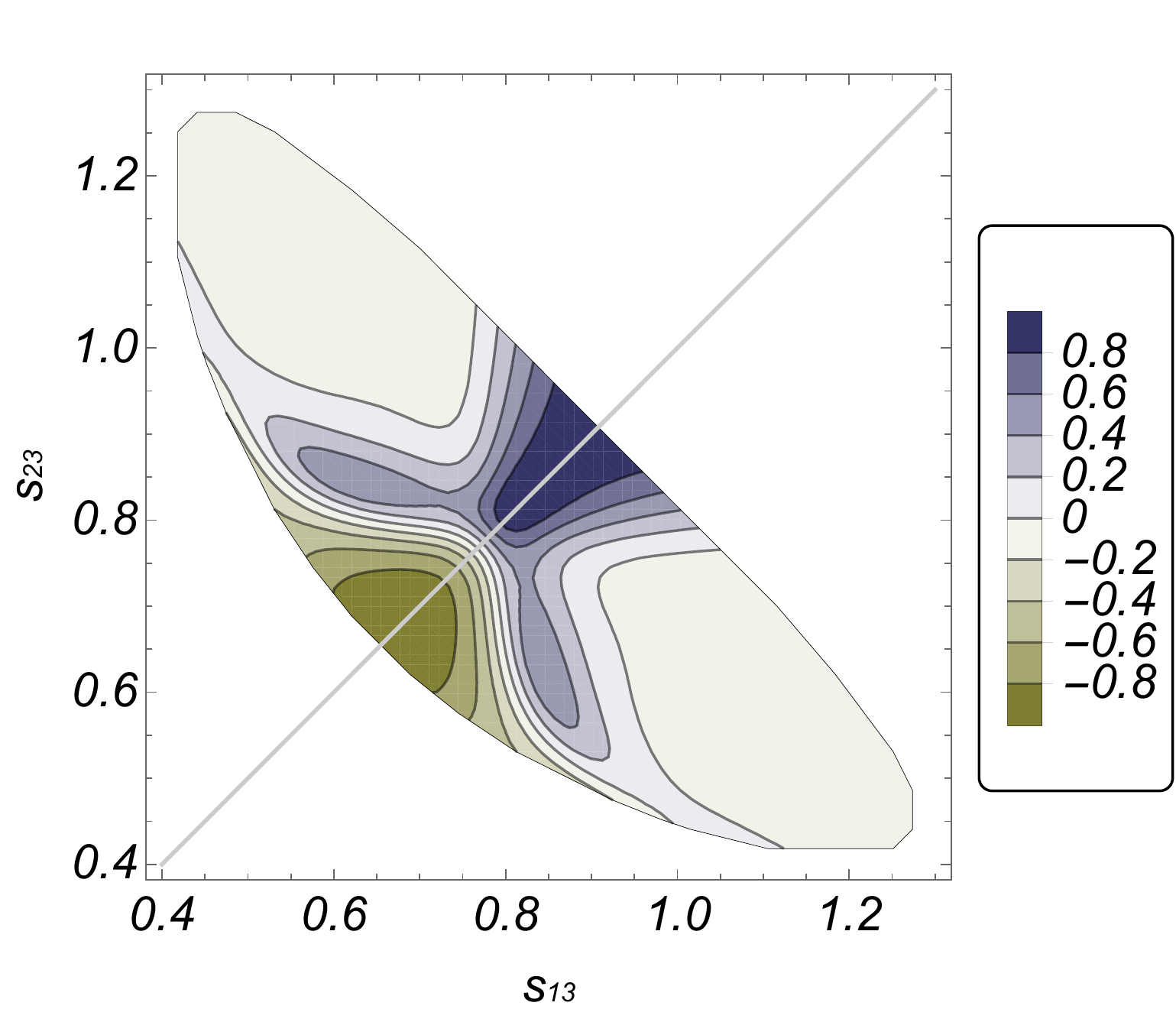}
\includegraphics[width = 0.45\linewidth]{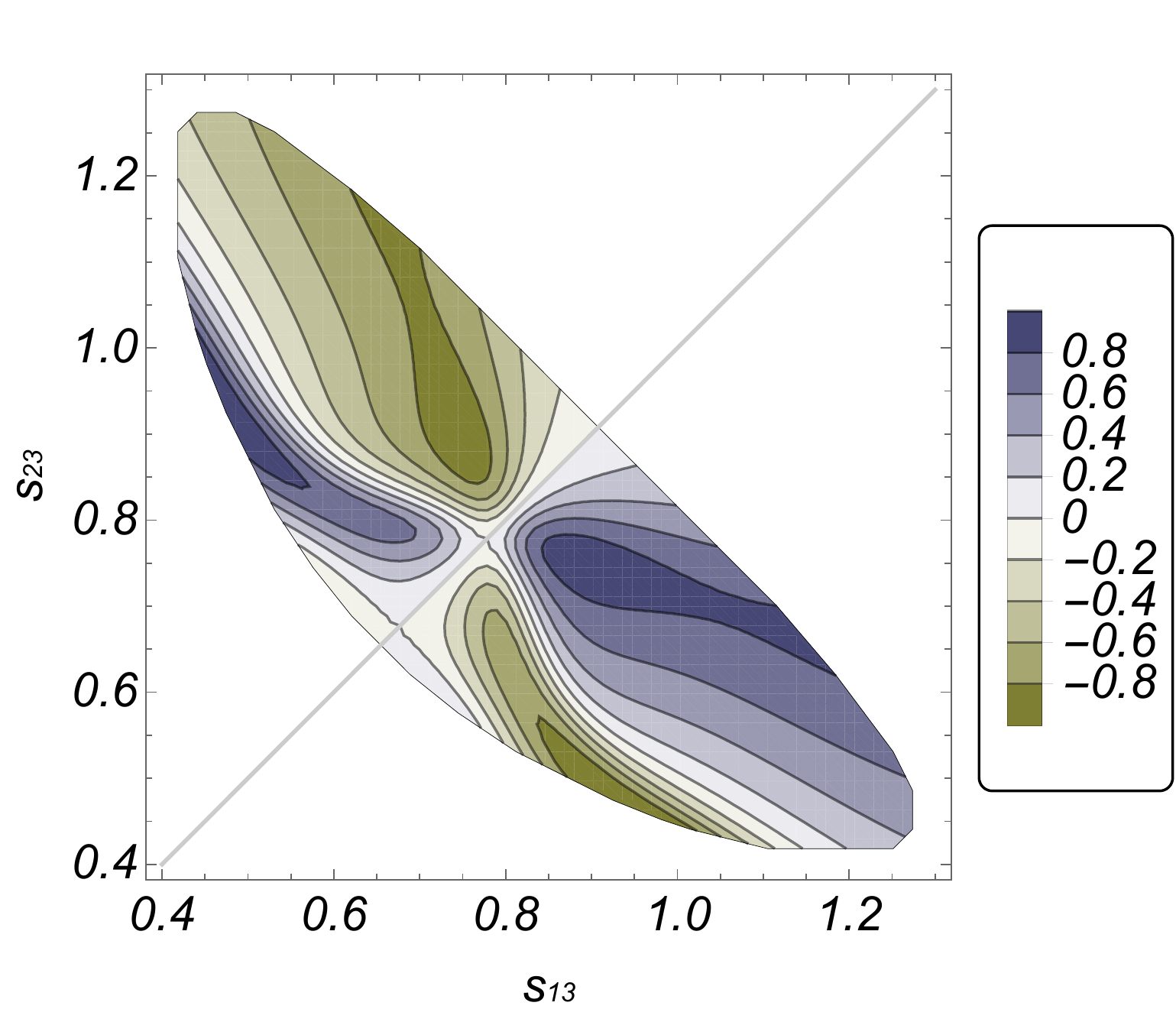}
\caption{\es{Normalised distributions of the real (left) and imaginary (right) parts of  \Eqref{eq:sec-5.2-ReIm_Amp} with   \mbox{$K_1(1270) \to \pip\pim\KS$} decay amplitudes, including all the intermediate resonances. } 
The axes correspond to $s_{13} = m^2_{\pip\KS}$ and $s_{23} = m^2_{\pim\KS}$ in \gevcccc. A similar behaviour is observed for all kaonic resonances.}
\label{fig:AllReIm}
\end{center}
\end{figure}
\es{Similarly to \secref{sec:lowStatStrategy}, the hadronic parameters, ${\rm a}^I$ and ${\rm b}^I$, need to be obtained from an amplitude analysis of $\Bp \to \Kresp \g \to \Kp \pim \pip \g$ decays.}
The partition scheme of the Dalitz plane must be optimised as a function of the amplitude content in the different regions and the available data sample.
From the anti-symmetric relation shown in \Eqref{eq:AmpRhoRelations}, it follows that the integrals of \es{ \Eqref{eq:sec-5.2-ReIm_Amp} with the \rhoKs amplitude} are real and independent of the integration region, with the values
\begin{equation}
	{\rm a}^{\delta\dalitz}_{\rhoKs}  = -\frac{1}{2}  \quad \textrm{ and } \quad  {\rm b}^{\delta\dalitz}_{\rhoKs} = 0.
\end{equation}
On the contrary, as shown in \figref{fig:KstReIm}, the real and imaginary parts of \es{\Eqref{eq:sec-5.2-ReIm_Amp} with the $\Kst\pi$ amplitude} vary as a function of the Dalitz-plane position. 
Furthermore, it clearly appears that the real (imaginary) part of the $\Kst\pi$ amplitude exhibits a symmetric (anti-symmetric) distribution with respect to the Dalitz plane bisector.
Similar behaviour is observed for the $\swave\pi$ amplitude.
As shown in \figref{fig:AllReIm}, when including the amplitudes of all the intermediate states, these symmetry properties with respect to the Dalitz plane bisector remain.

%% file: prospects.tex
\section{Constraints on new physics and future prospects}\label{sec:interpretation}

Finally, the constraints on $\cprime/c$, which can be obtained from the time-dependent measurement of $\Bz \to \pip\pim \KS \g$ decays, are discussed.
The common name of $\cprime/c$ is \CSpoverCS; we use this notation hereafter. 

\begin{figure}[b]
\begin{center}
\includegraphics[width = 0.45\linewidth]{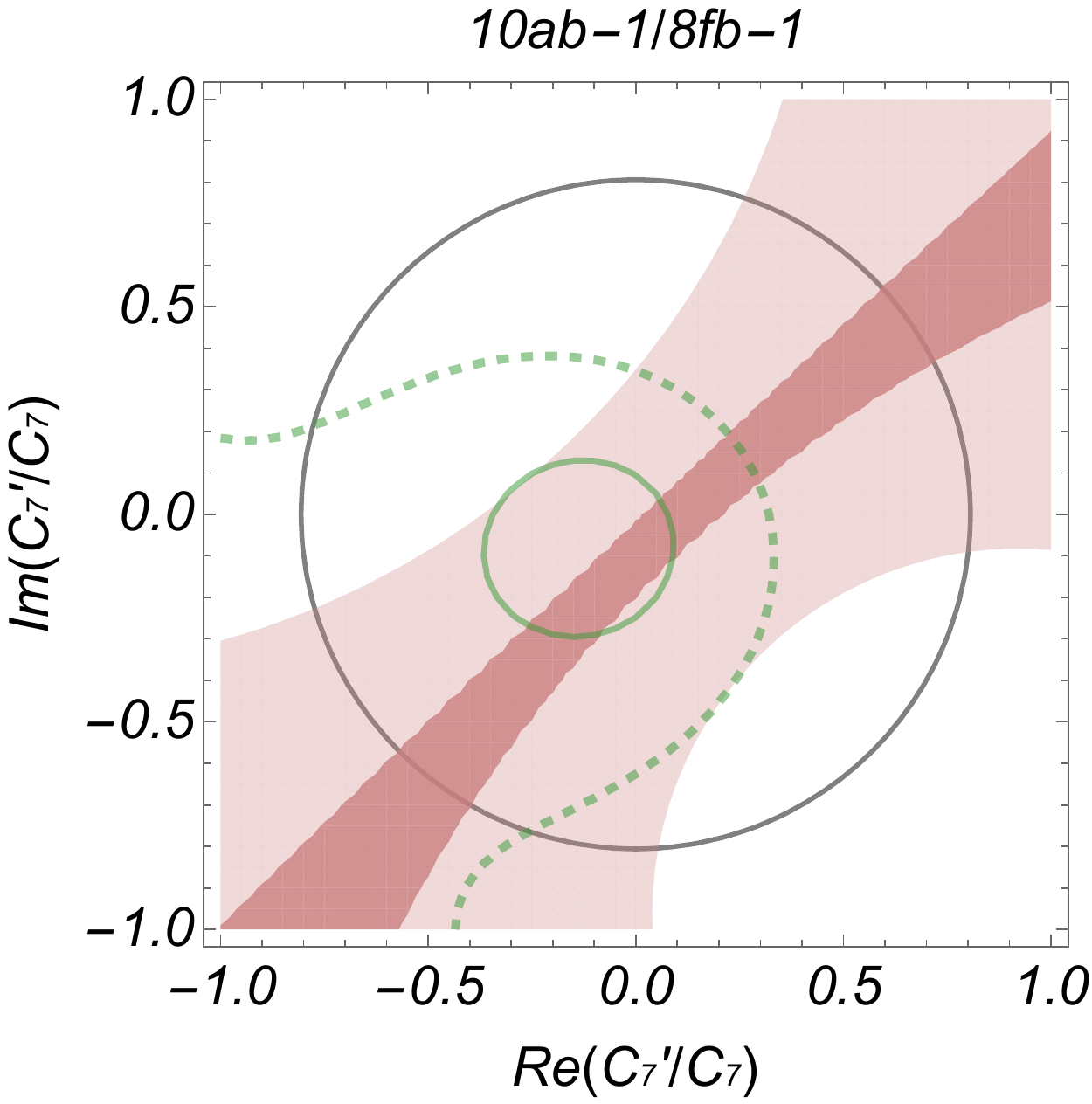}\ \ \ \ \ 
\includegraphics[width = 0.45\linewidth]{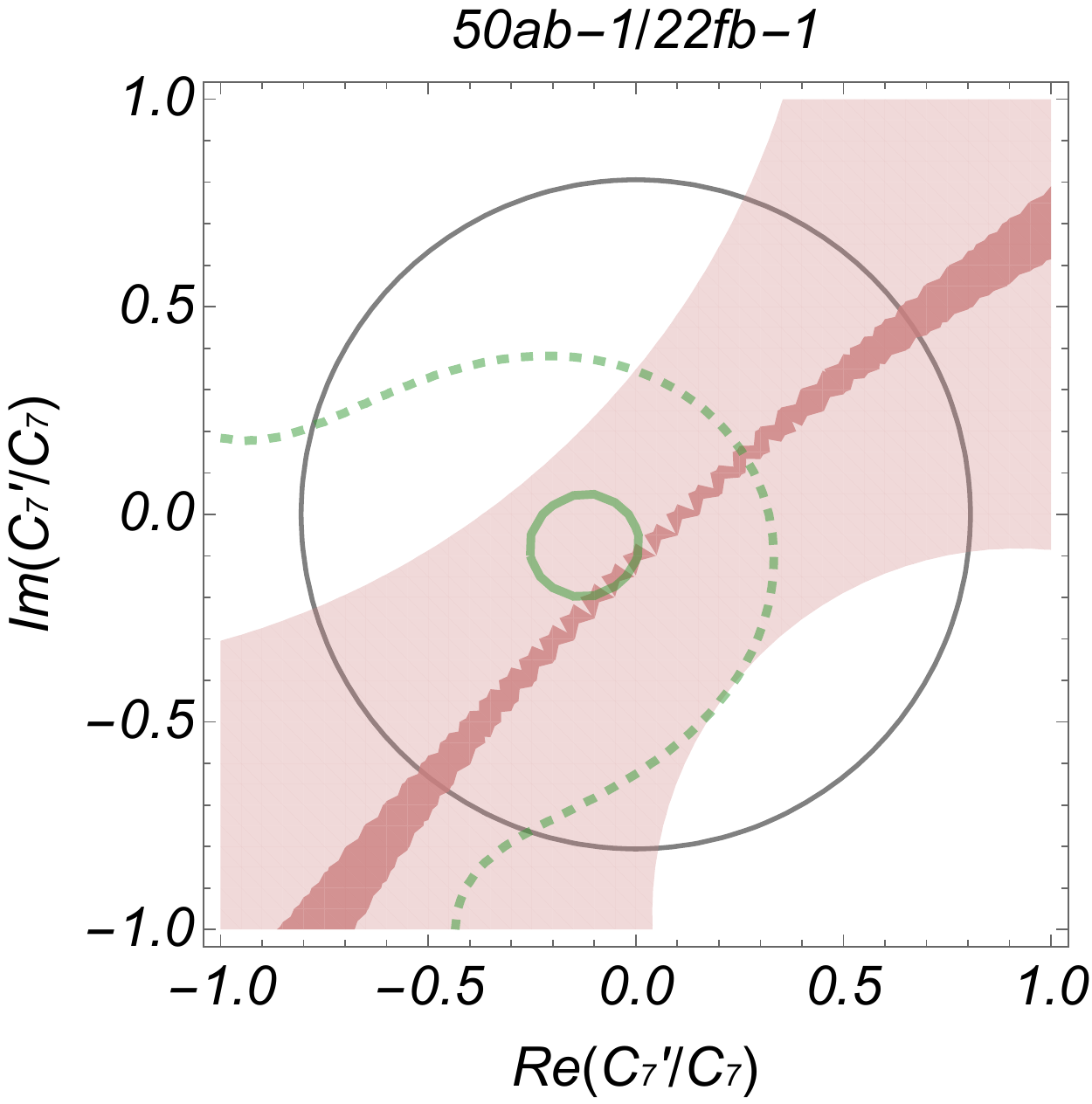}
\caption{
Constraints on \Real{\CSpoverCS} and \Imag{\CSpoverCS} at the three standard-deviations level.
The light red region is the constraint obtained from the current measurement of time-dependent \CP asymmetry in $\Bz \to \piz \KS \g$ decays,  $\SKspiz=-0.15\pm0.20$~\cite{Amhis:2016xyh}, overlaid with the constraint (dark red) obtained from the expected precision at \superbelle with integrated luminosities of $10\ab$ (left) and $50\ab$ (right). 
The dashed green contour is the constraint obtained from the angular coefficients of $B\to K^*e^+e^-$ decays at $q^2\to 0$, measured by LHCb~\cite{Aaij:2015dea}, $A_T^{(2)}=-0.23\pm 0.24$ and $A_T^{\rm im}=0.14\pm 0.23$, overlaid with the contraint (full green line) obtained from the expected precision at LHCb Run II (8 fb$^{-1}$) and Run III (22 fb$^{-1}$).
The grey circular contour is the constraint obtained from the branching fraction measurement of the inclusive $B\to X_s\gamma$ processes with ${\cal B}(B\to X_s\gamma)^{E_\gamma>1.6 {\rm GeV}}_{\rm exp}=3.27\pm 0.14$~\cite{Amhis:2016xyh} and  ${\cal B}(B\to X_s\gamma)^{E_\gamma>1.6 {\rm GeV}}_{\rm th}=3.36\pm 0.23$~\cite{Misiak:2015xwa}.}
\label{fig:CspoverCS_SKSPiz}
\end{center}
\end{figure}
Currently, the most stringent constraints on \CSpoverCS are obtained from the time-dependent \CP asymmetry in $B\to K_S\pi^0\gamma$ decays, the angular coefficients of $B\to K^*e^+e^-$ decays at $q^2\to 0$, $A_T^{(2)}$ and $A_T^{Im}$, and the branching fraction of the inclusive $B\to X_s\gamma$ process.
In \figref{fig:CspoverCS_SKSPiz}, we show the constraints on \Real{\CSpoverCS} and \Imag{\CSpoverCS} at the three standard-deviations level, obtained from the available measurements. 
\es{The figure also shows the constraints that can be obtained with the expected precision at \superbelle with datasets of $10\ab$ (foreseen by the year $\sim$2023) and $50\ab$ ($\sim$2027) and at LHCb with datasets of $8\fb$ (current) and $22\fb$ ($\sim$2023).}
The expected constraints are obtained by using the current central values of the observables, and assuming the measurements to be  limited by the statistical uncertainties. 
As it is well known, \SKspiz provides a precise determination of \CSpoverCS, even though it cannot disentangle its real part from its imaginary part.
\begin{figure}[b]
\begin{center}
\includegraphics[width = 0.45\linewidth]{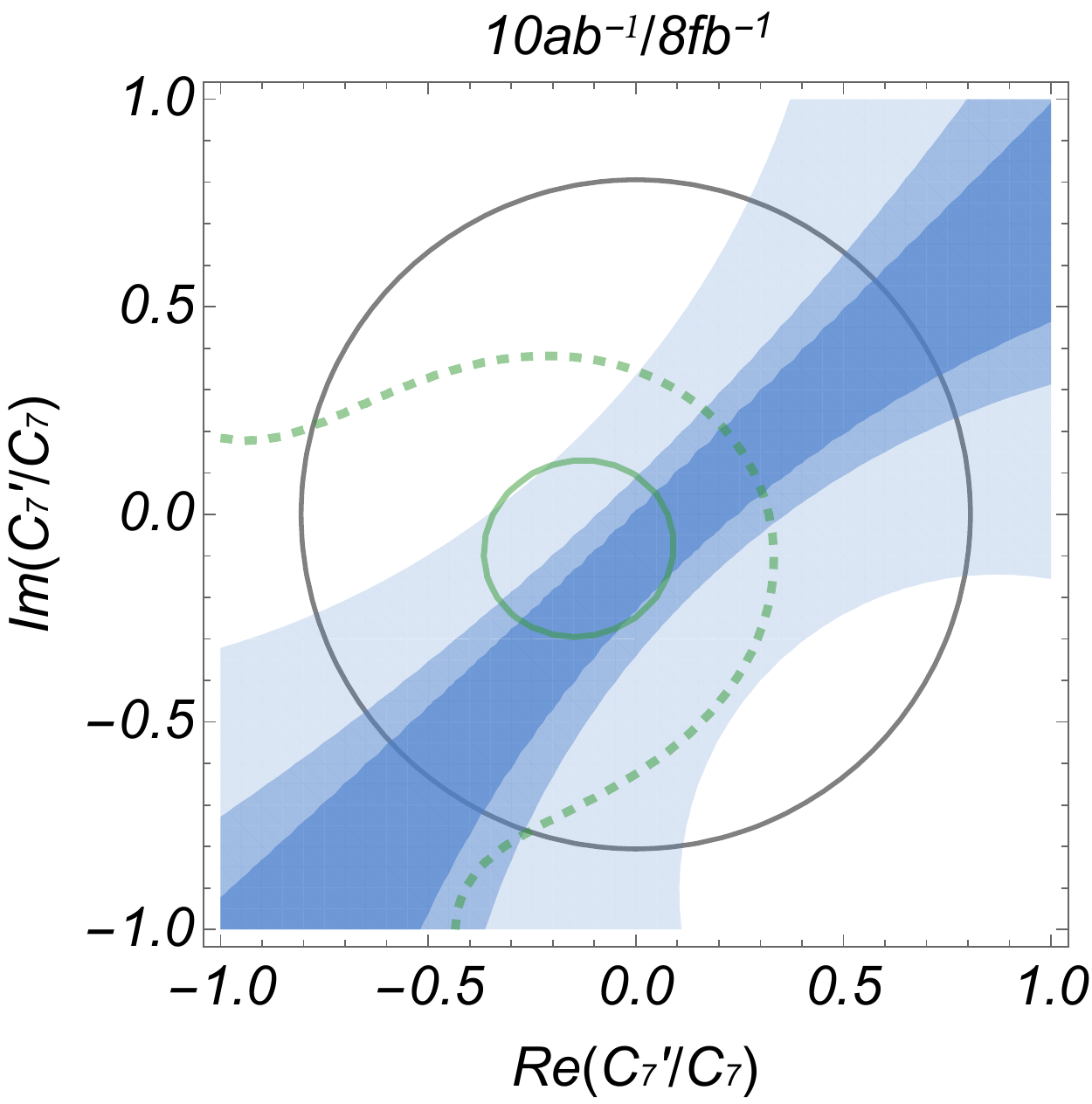}\ \ \ \ \ 
\includegraphics[width = 0.45\linewidth]{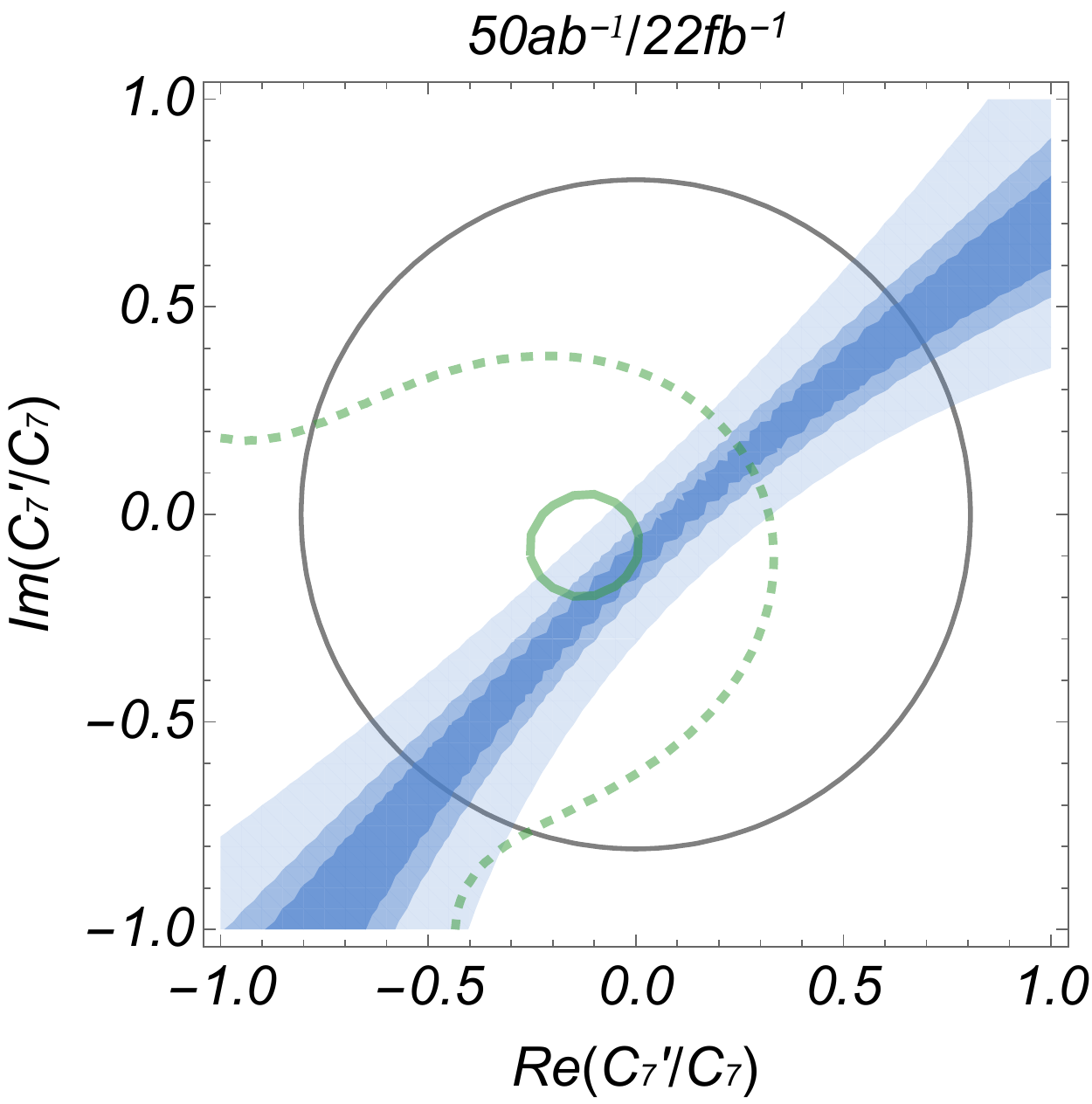}
\caption{Prospects for constraining, at the three standard-deviations level, \Real{\CSpoverCS} and \Imag{\CSpoverCS} from a measurement of \SKspipi, integrating over the whole Dalitz plane. 
The colour changes from dark blue to light blue for decreasing values of the dilution factor ${\D=\{1,0.6,0.3\}}$. 
For each of these, a central value of \SKspipi is chosen arbitrarily to be, respectively, $\SKspipi=\{0.15, 0.09, 0.05 \}$, and the current experimental uncertainty is scaled by the increase of integrated luminosity.
The grey and green contours are described in  \figref{fig:CspoverCS_SKSPiz}.}
\label{fig:CSpoverCS_Sp}
\end{center}
\end{figure}

Next, let us show the expected constraint from \SKspipi, integrating over the whole Dalitz plane, as described in \secref{sec:lowStatStrategy}.
The determination of \Real{\CSpoverCS} and \Imag{\CSpoverCS} via \SKspipi depends on the value of the dilution factor, hence on the amplitudes of intermediate states and on the integration region. 
In \figref{fig:CSpoverCS_Sp}, we show the constraints for different values of the dilution factor, ${\D=\{1,0.6,0.3\}}$. 
For the sake of demonstration, the central values are arbitrarily chosen as $\SKspipi=\{0.15, 0.09, 0.05\}$, to facilitate the comparison with the constraints on \CSpoverCS from \SKspiz as shown in \figref{fig:CspoverCS_SKSPiz}. 
The experimental uncertainties are obtained by scaling the statistical uncertainty, $\sigma(\SKspipi)=0.25$, from the latest \babar analysis~\cite{Sanchez:2015pxu} with a dataset of $\sim$ $0.5\ab$, to the precision expected with integrated luminosities of $10\ab$ and $50\ab$ at \superbelle. 
The uncertainty on the dilution factor, $\sigma(\D)=0.18$~\cite{Sanchez:2015pxu}, is scaled in the same way. 
Note that $\D=1$ corresponds to the case where the intermediate mode $\Bz \to \rhoz \KS \g$ dominates and the effect of other resonances is negligible. 
The contributions of the other intermediate modes, i.e. $\Bz \to \Kst\pi\gamma$ and $\Bz \to \swave\pi\gamma $, decrease the value of \D, resulting in looser constraints on \CSpoverCS. 
This result shows the importance of carefully optimising the phase space region in the measurement of \SKspipi to ensure the best interplay between the dilution factor and the number of events.
\begin{figure}[t]
\begin{center}
\includegraphics[width = 0.45\linewidth]{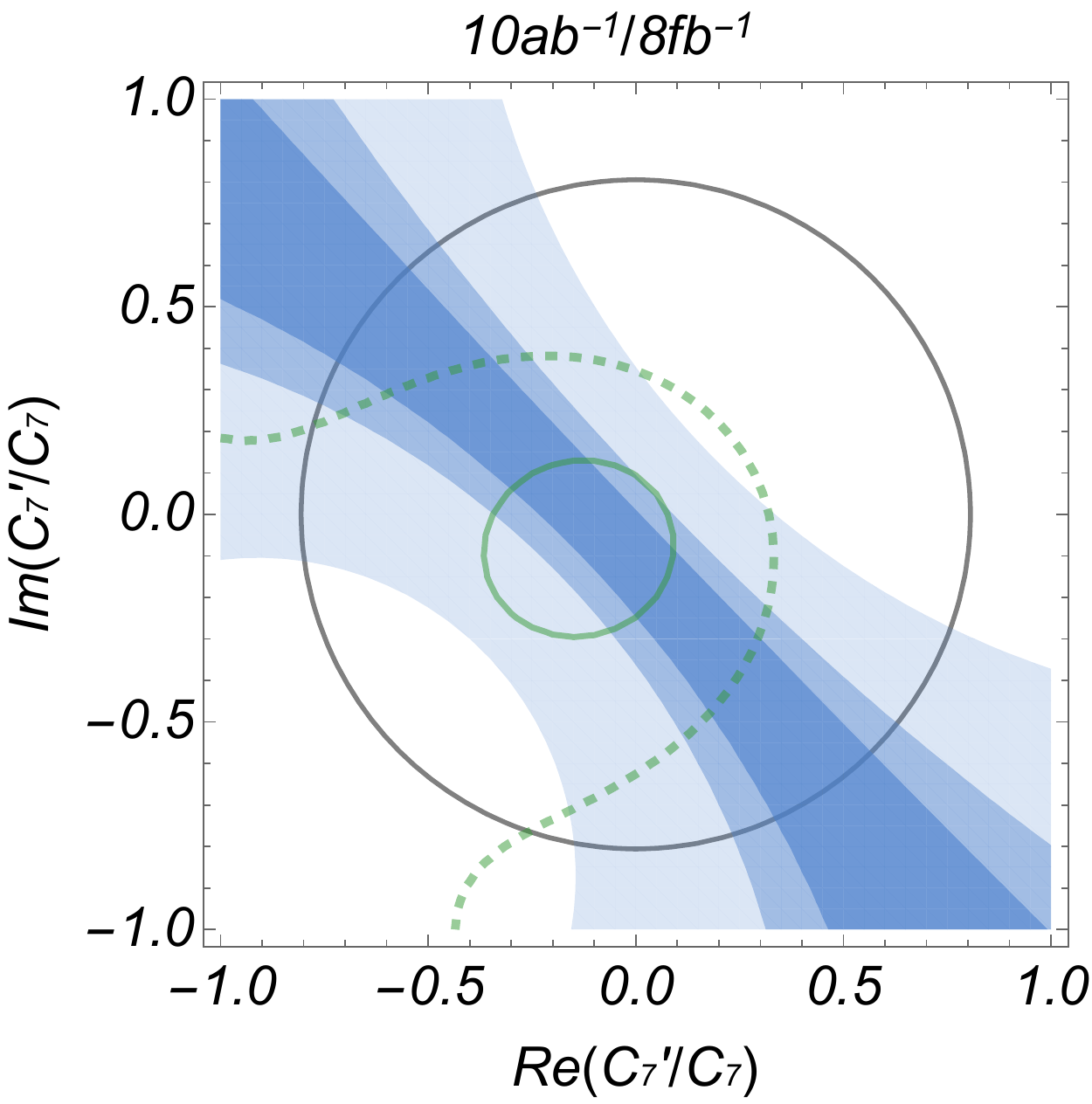}\ \ \ \ \ 
\includegraphics[width = 0.45\linewidth]{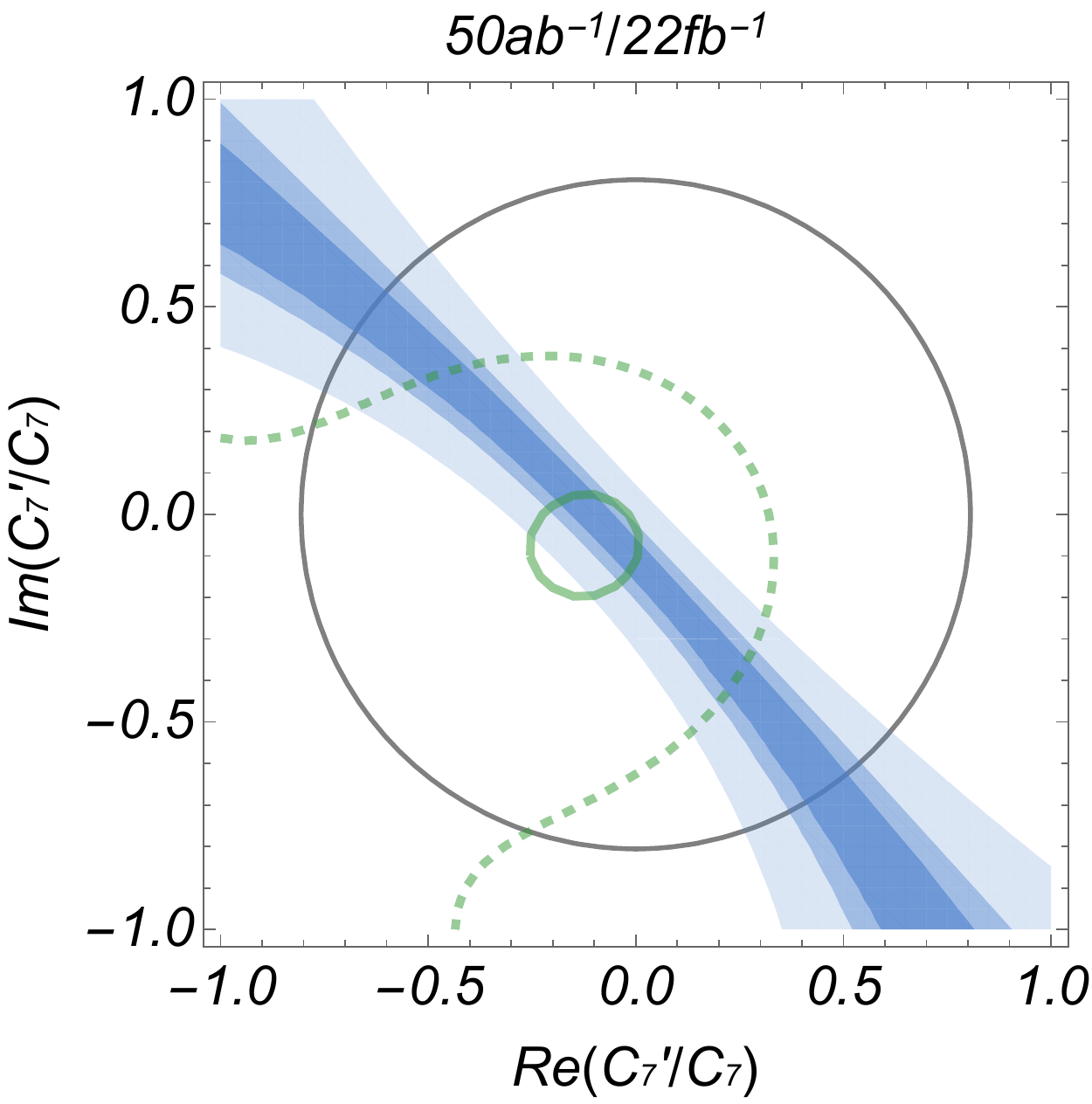}
\caption{
Prospects for constraining, at the three standard-deviations level, \Real{\CSpoverCS} and \Imag{\CSpoverCS} from the proposed observable \SM.
The colour changes from dark blue to light blue for increasing values of
the hadronic parameter ${\rm b}^I=\{-0.5,-0.3,-0.15\}$. 
The corresponding values of $\SM=\{0.30, 0.18, 0.10\}$  are chosen to match those used for \SKspipi in \figref{fig:CSpoverCS_Sp}, with similar uncertainties.
The uncertainty on the hadronic parameter ${\rm b}^I$ is accounted for, assuming $\sigma({\rm b}^I) = \sigma(\D)/\sqrt{2}$. 
All uncertainties are further scaled according to the increase of integrated luminosity.
The grey and green contours are described in \figref{fig:CspoverCS_SKSPiz}.}
\label{fig:CSpoverCS_Sm}
\end{center}
\end{figure}

Let us now consider the proposed observable \SM of \Eqref{eq:SKspipiDalitzIb}, representing the difference of the time-dependent \CP asymmetries measured in two regions of the Dalitz plane. 
The observable \SP in \Eqref{eq:SKspipiDalitzI} yields a similar constraint as that from the integrated analysis, which is shown in \figref{fig:CSpoverCS_Sp}. 
On the contrary, \SM leads to a different kind of constraint; an example is shown
in~\figref{fig:CSpoverCS_Sm}, with the
central values $\SM = \{0.30, 0.18, 0.10\}$, chosen to match those of \SKspipi in \figref{fig:CSpoverCS_Sp}, and with the hadronic parameter values ${\rm b}^{I}=\{-0.5,-0.3,-0.15\}$. 
The uncertainties on \SM are obtained assuming 
\mbox{$\sigma(S^{I}) = \sigma(S^{\overline{I}}) = \sqrt{n}\sigma(\SKspipi)$}, where $n=2$ corresponds to the number of Dalitz-plane regions, and neglecting the correlations between $S^{I}$ and $S^{\overline{I}}$.
The uncertainties on ${\rm b}^{I}$ are obtained assuming \mbox{$\sigma({\rm b}^I) = \sigma({\rm a}^I) = \sigma(\D)/\sqrt{n}$}.
The relation \mbox{$\sigma({\rm b}^I) = \sigma({\rm a}^I)$} is obtained assuming that the uncertainty on the magnitude of the hadronic decay is similar to that on the arc-length corresponding to the phase difference.

It is remarkable that the constraint on \CSpoverCS obtained from \SM is orthogonal to that from \SP. 
From \Eqsref{eq:SKspipiDalitzI}{eq:SKspipiDalitzIb}, it is clear that this orthogonality does not depend on the values chosen for this demonstration.
Thus, by combining the two observables it is possible to obtain stringent constraints on both the real and imaginary parts of \CSpoverCS. 
On the other hand, it is not possible to obtain such contraints from a time-dependent analysis of $\Bz \to \piz \KS \g$ decays, where \SM and \SP are not defined.
\begin{figure}[t]
\begin{center}
\includegraphics[width = 0.45\linewidth]{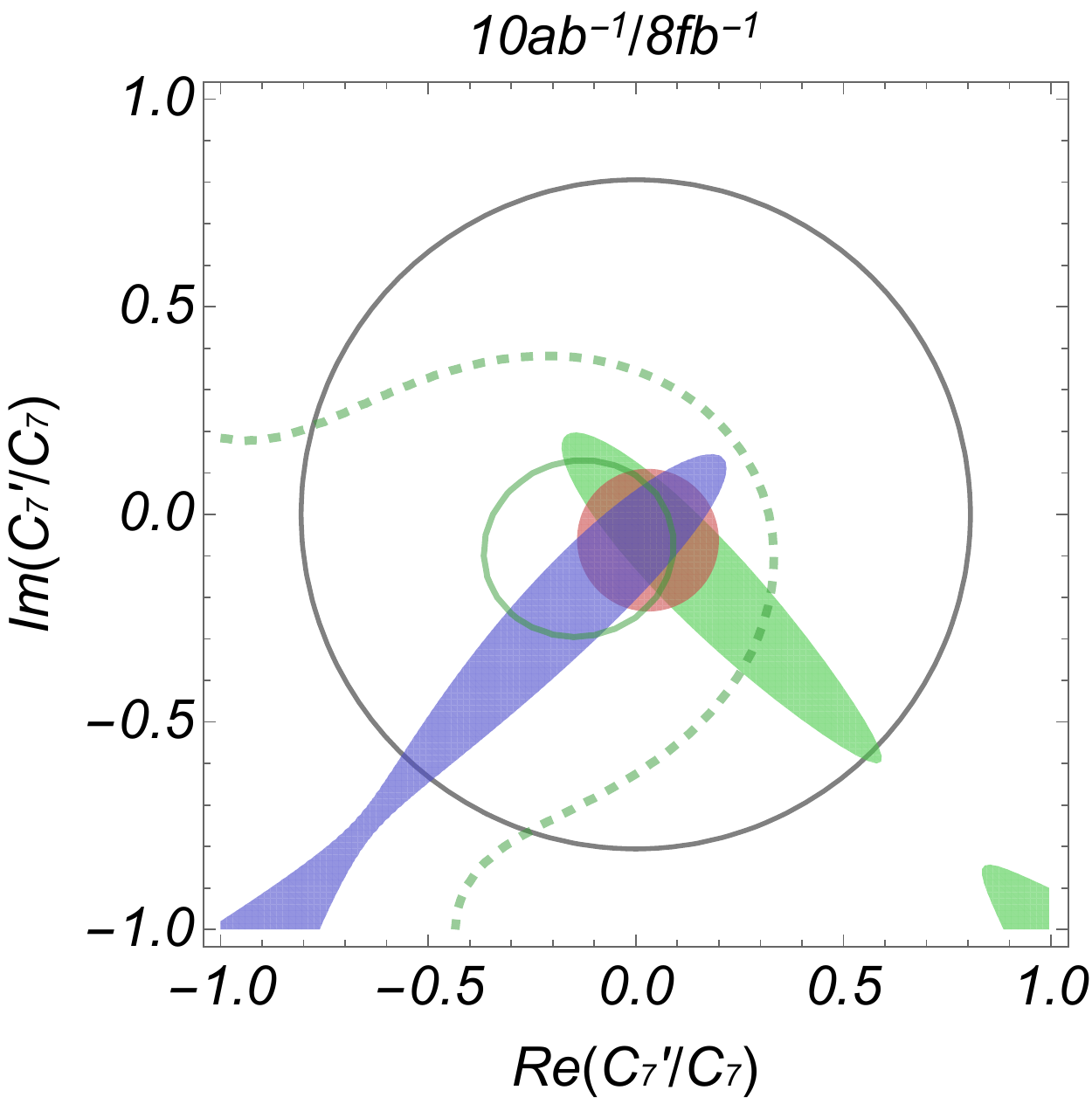}\ \ \ \ \ 
\includegraphics[width = 0.45\linewidth]{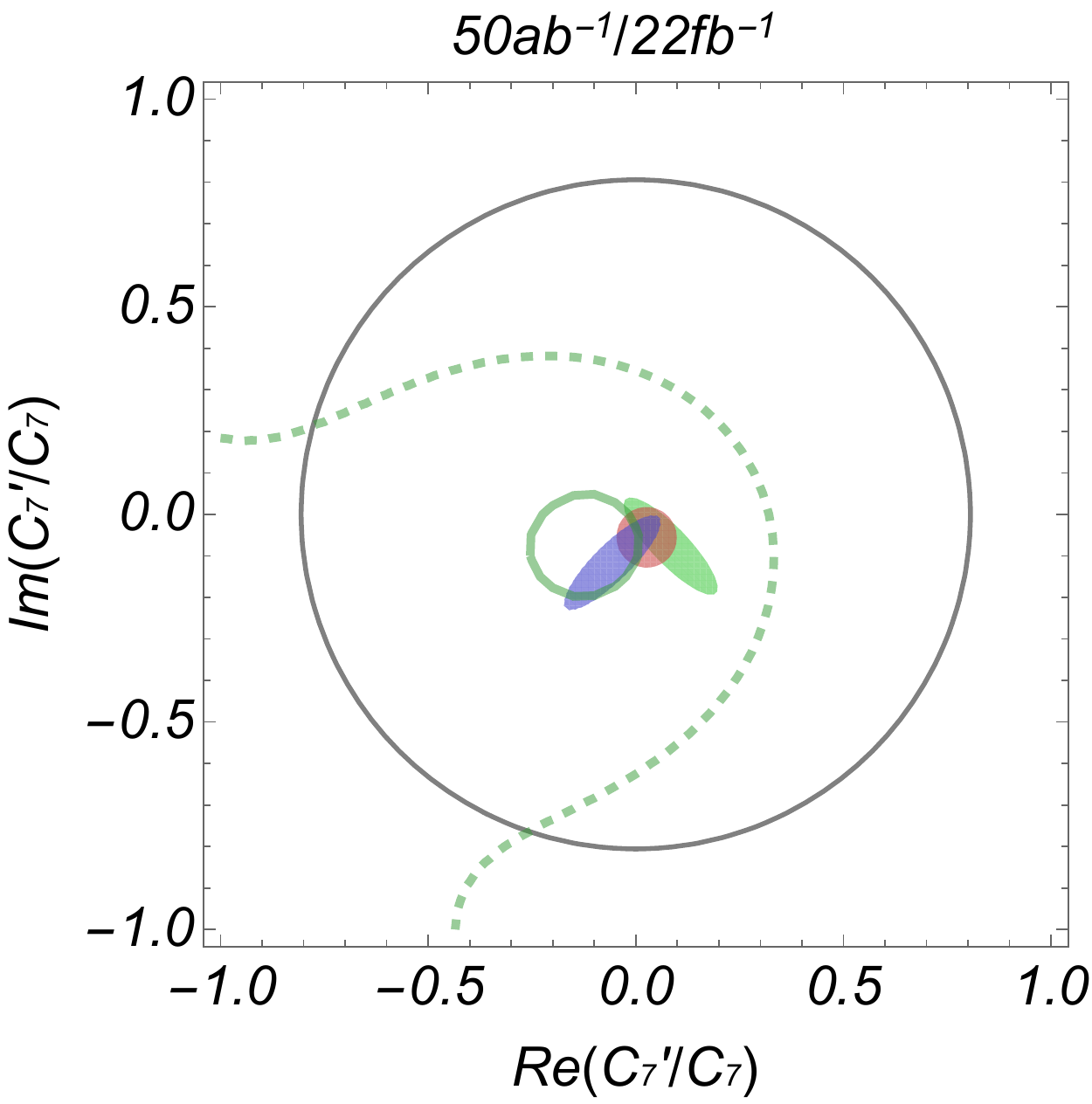}
\caption{
Prospects for constraints on \Real{\CSpoverCS} and \Imag{\CSpoverCS}, at the three standard-deviations level, obtained from the two observables \SM and \SP.
The central value of \SM is chosen arbitrarily, while its uncertainty is estimated using results from Ref.~\cite{Sanchez:2015pxu}. 
Three sets of central values, yielding different contraints on \Real{\CSpoverCS} and \Imag{\CSpoverCS}, are chosen: 
$\{\SP, \SM, {\rm a}^I, {\rm b}^I\}=\{0.17, 0.13, -0.5, -0.15\}$ (blue), $\{0.13, 0.04, -0.3, -0.3\}$ (red) and $\{0.13, -0.03, -0.15, -0.5\}$ (green).
The uncertainties are taken as \mbox{$\sigma(S^{I}) = \sigma(S^{\overline{I}}) = \sqrt{n}\sigma(\SKspipi)$}, and \mbox{$\sigma({\rm a}^I) = \sigma({\rm b}^I) = \sigma(\D)/\sqrt{n}$}, where $n=2$ corresponds to the number of Dalitz-plane regions.
All the uncertainties are further scaled according to the increase of integrated luminosity. 
The grey and green contours are described in \figref{fig:CspoverCS_SKSPiz}. }
\label{fig:CSpoverCS_Combined}
\end{center}
\end{figure}

We finally show, in \figref{fig:CSpoverCS_Combined}, an example of the combined constraints from \Eqsref{eq:SKspipiDalitzI}{eq:SKspipiDalitzIb}. 
As before, the central values of \SP, \SM, ${\rm a}^{I}$ and $ {\rm b}^{I}$ are chosen arbitrarily,
and the uncertainties are estimated from the \babar measurements of \SKspipi and \D~\cite{Sanchez:2015pxu}. 
The sets of central values used in \figref{fig:CSpoverCS_Combined} are $\{\SP, \SM, {\rm a}^I, {\rm b}^I\}=\{0.17, 0.13, -0.5, -0.15\}$, $\{0.13, 0.04, -0.3, -0.3\}$ and $\{0.13, -0.03, -0.15, -0.5\}$.
Even though the obtained constraints depend on the hadronic parameters, it is clear that combining the information from \SP and \SM measured in $\Bz \to \pip\pim \KS \g$ decays allows to independently constrain both \Real{\CSpoverCS} and \Imag{\CSpoverCS}. 

%% file: conclusion.tex
\section{Conclusion}
\label{sec:conclusion}

In this paper, we derive the formula for the time-dependent \CP asymmetry of $\Bz \to \Kres \g \to [\rhoKs, \KstPi, \KappaPi]\g \to \pip \pim \KS \g$; it is the first time that this formula is derived including all these intermediate states. 
As it turns out, the formula is the same for all \Kres states with $J^{P}=(1^+, 1^-, 2^+)$.
This allows to extract the time-dependent \CP asymmetry \SKsrho by measuring the phase-space integrated \SKspipi and the dilution factor \D.
The constraint from this measurement on the \CSpoverCS complex plane is similar to that obtained from the measurement of \SKspiz; it corresponds to a diagonal band.
The dilution factor can be obtained from the charged decay mode $\Bp \to \pip \pim \Kp \g$, which benefits from a higher branching fraction and a better detection efficiency compared to the neutral decay mode.
In particular, the LHCb experiment is currently in the best position to provide additional information on \D.  

We also show that performing a time-dependent amplitude analysis of $\Bz \to \pip \pim \KS \g$ decays gives access to a new observable, \SM, which allows, when combined with \SP, to obtain stringent constraints on both the real and imaginary parts of \CSpoverCS.
This is the main result of this paper.
Such an analysis is not currently feasible due to the limited size of the available data samples, whereas it will become accessible with the dataset expected from the \superbelle experiment.
We present prospects for the determination of \CSpoverCS from a time-dependent analysis of $\Bz \to \pip \pim \KS \g$ decays at \superbelle, considering two approaches: a phase-space integrated analysis and an amplitude analysis using information from the $\KS\pip\pim$ Dalitz-plane. 

The analysis of $\Bz \to \pip \pim \KS \g$ decays should provide stringent constraints on the photon polarisation in the upcoming years. 
In particular, the constraints on \CSpoverCS from this measurement are complementary to those obtained from the time-dependent \CP asymmetry of $\Bz \to \piz \KS \g$ and the angular analysis of $B \to \Kst e^+e^-$ at low $q^2$.

%% file: acknowledgements.tex
\section*{Acknowledgements}

We would like to thank Fran\c cois Le Diberder, David London and Michael D. Sokoloff for fruitful discussions concerning this paper and their valuable comments and advice. This work was supported, in part, by the U.S. National Science Foundation.

%% file: appendix_S-KsPiz_S-Phi.tex
\appendix
\section{\boldmath{The time-dependent \CP asymmetry in $\Bz \to \Kstz \g \to \piz \KS\g$  and $\Bs \to \phi \g \to \Kp \Km \g$}}
\label{App:CompOtherModes}

For comparison, we include the \CP formulae for the \mbox{$\Bz \to \Kstz \g \to \piz \KS\g$} and \mbox{$\Bs \to \phi \g \to \Kp \Km \g$} in this appendix. 

Let us obtain the amplitude relations, as done in \secref{sec:amplitudes}.
The relations between the right and left handed amplitudes in \Eqref{eq:2-19} hold also in the cases of 
\mbox{$\Bz \to \Kstz \g$} and \mbox{$\Bs \to \phi \g$} decays.
For the strong amplitude \mbox{$\Kstz \to \piz \KS$}, the \C transformation leads to
\begin{eqnarray} 
\AmpStrong^{\piz\KS}&=&
      \braket{ \KS(p_1)\piz(p_2) }{ \Hsprime }{ \Kst }  \nonumber \\
  &=& \braket{ \KS(p_1)\piz(p_2) }{ \Hsprime }{ \Kstb } ,\nonumber  
\end{eqnarray}
where we assigned $\C \ket{\Kstz} = - \ket{\Kstzb}$ for consistency. 
For $\phi \to \Kp\Km$ 
\begin{eqnarray} 
\AmpStrong^{\prime \Kp\Km}&=&
        \braket{\Kp(p_1)\Km(p_2)}{\Hsprime}{\phi} \nonumber \\
  &=& - \braket{\Km(p_1)\Kp(p_2)}{\Hsprime}{\phi} \nonumber \\ \nonumber 
  &=&   \braket{\Km(p_2)\Kp(p_1)}{\Hsprime}{\phi}, \nonumber  
\end{eqnarray}
where the last line is explained by the fact that the $\phi$ decays trough a p-wave. 
For the parity transformation, \Eqref{eq:v3-30} holds here as well. 
Using these relations, for  $\Bz \to \Kst \g$ and $\Bs \to \phi \g$ decays we find 
\begin{eqnarray}
\AmpTotL^{*\piz\KS}\AmpTotBarL^{\piz\KS} &=&\left(\covercprimestar\right) \ModSq{\AmpTotL^{\piz\KS}}, \\
\AmpTotL^{*\Kp\Km}\AmpTotBarL^{\Kp\Km} &=&\left(\covercprimestar\right) \ModSq{\AmpTotL^{\Kp\Km}} , \end{eqnarray}
and \Eqref{eq:4-3} holds here as well. 
As a result, we find for  $\Bz \to \Kst \g$ 
\begin{equation}
\CKspiz=0, \quad \SKspiz =
\frac{2\Imag{\qoverp \Wc \cprime}}{\ModSq{\Wc}+\ModSq{\cprime}}.
\end{equation}

\noindent For the \mbox{$\Bs \to \phi \g \to \Kp \Km\g$} decay, \Eqref{eq:2-1} has an additional term due to the large $\Delta \Gamma_s$ with respect to $\Delta \Gamma_d$\cite{Amhis:2016xyh} such that
\begin{equation}
\frac{\overline{\Gamma}(t)-\Gamma(t)}{\overline{\Gamma}(t)+\Gamma(t)}\equiv 
\frac{\mathcal{S}_{\Kp\Km} \sin (\dm t)-\mathcal{C}_{\Kp\Km}\cos (\dm t)}{\cosh\frac{\Delta\Gamma_st}{2}-\mathcal{A}^{\Delta}_{\Kp\Km\gamma}\sinh\frac{\Delta\Gamma_st}{2}},
\end{equation}
where $\mathcal{A}^{\Delta}_{\Kp\Km\gamma}$ is given by 
\begin{equation}
\mathcal{A}^{\Delta}_{\Kp\Km\gamma} =\frac{2\Real{\qoverp\sumLR{\AmpTotConj \AmpTotBar}}}{\sumLR{\ModSq{\AmpTotBar}+\ModSq{\AmpTot}}}.
\end{equation}
Then, we find 
\begin{equation}
\mathcal{C}_{\Kp\Km\gamma}=0, \quad \mathcal{S}_{\Kp\Km\gamma}=
\frac{2\Imag{\frac{q_s}{p_s}
c_d c_d^{\prime}}}{\ModSq{c_d}+\ModSq{c_d^{\prime}}},
\quad 
\mathcal{A}^{\Delta}_{\Kp\Km\gamma}=\frac{2\Real{\frac{q_s}{p_s}c_d c_d^{\prime}}}{\ModSq{c_d}+\ModSq{c_d^{\prime}}},
\end{equation}
where $q_{s}/p_s$ indicates the $\Bs-\Bsb$ mixing phase and $c_d$ indicates the coefficient of the \btodg transition amplitude. Integration over the whole phase space is implicit.  
The first measurement of $\mathcal{A}^{\Delta}_{\Kp\Km\gamma}$ was obtained by the LHCb collaboration~\cite{Aaij:2016ofv}. 
This result was recently superseded by an updated analysis also including the first measurements of $\mathcal{S}_{\Kp\Km\gamma}$ and $\mathcal{C}_{\Kp\Km\gamma}$~\cite{Aaij:2019pnd}.